\documentclass[11pt,pra,showpacs,superscriptaddress,floatfix]{revtex4-2}
\usepackage{graphicx}
\usepackage{amssymb}
\usepackage{amsmath}
\usepackage{amsthm}
\usepackage{bm}
\usepackage[colorlinks,
linkcolor=blue,      
anchorcolor=blue, 
citecolor=purple]{hyperref}
\usepackage{color,xcolor}
\usepackage{subfigure}
\usepackage{algpseudocode}
\usepackage{soul}
\usepackage{algorithm}
\usepackage{algorithmicx}
\usepackage{lineno}
\usepackage{soul,xcolor}
\usepackage{color}

\usepackage{graphicx}
\usepackage{dcolumn}
\usepackage{bm}
\usepackage{graphicx}
\usepackage{textcomp,mathcomp}
\usepackage{enumerate}

\begin{document}

\title{Real-time imaging through dynamic scattering media enabled by fixed optical modulations}

\author{Yuegang Li}
 \affiliation{State Key Laboratory of Photonics and Communications, Institute for Quantum Sensing and Information Processing, Shanghai Jiao Tong University, Shanghai 200240, P.R. China}

 \author{Junjie Wang}\affiliation{State Key Laboratory of Photonics and Communications, Institute for Quantum Sensing and Information Processing, Shanghai Jiao Tong University, Shanghai 200240, P.R. China}

 \author{Tailong Xiao}%
\email{tailong\_shaw@sjtu.edu.cn}
\affiliation{State Key Laboratory of Photonics and Communications, Institute for Quantum Sensing and Information Processing, Shanghai Jiao Tong University, Shanghai 200240, P.R.  China}%
 \affiliation{Hefei National Laboratory, Hefei, 230088, P.R. China}
\affiliation{Shanghai Research Center for Quantum Sciences, Shanghai, 201315, P.R. China}

 \author{Ze Zheng}\affiliation{State Key Laboratory of Photonics and Communications, Institute for Quantum Sensing and Information Processing, Shanghai Jiao Tong University, Shanghai 200240, P.R. China}  

\author{Jingzheng Huang}
\affiliation{State Key Laboratory of Photonics and Communications, Institute for Quantum Sensing and Information Processing, Shanghai Jiao Tong University, Shanghai 200240, P.R. China}%
 \affiliation{Hefei National Laboratory, Hefei, 230088, P.R. China}
\affiliation{Shanghai Research Center for Quantum Sciences, Shanghai, 201315, P.R. China}

\author{Ming He}
\affiliation{AI Lab, Lenovo Research, Beijing 100094, P.R. China}
 
\author{Jianping Fan}
\affiliation{AI Lab, Lenovo Research, Beijing 100094, P.R. China}

\author{Guihua Zeng}
\email{ghzeng@sjtu.edu.cn}
\affiliation{State Key Laboratory of Photonics and Communications, Institute for Quantum Sensing and Information Processing, Shanghai Jiao Tong University, Shanghai 200240, P.R. China}%
 \affiliation{Hefei National Laboratory, Hefei, 230088, P.R. China}
\affiliation{Shanghai Research Center for Quantum Sciences, Shanghai, 201315, P.R. China}
 

\begin{abstract}

Dynamic scattering remains a significant challenge to the practical deployment of anti-scattering imaging. Existing methods, such as transmission matrix measurements, iterative wavefront shaping, and optical phase conjugation, depend on a quasi-static assumption, requiring the object and scattering medium to remain stable during a single imaging process to enable one-to-one compensation. However, image reconstruction becomes unattainable when this assumption is violated. Here, we propose a novel imaging strategy that counteracts time-dependent scattering perturbations through a fixed modulation module. This one-to-many compensation mechanism is realized via optical diffraction neural networks (ODNNs) trained on simulated datasets. For the first time, we reveal that its feasibility stems from the optical shower-curtain effect, and its effectiveness typically  within 1–2 transport mean free paths.   Our approach is not only immune to speckle decorrelation but also leverages it to enhance image quality. ODNNs generalize effectively to real-world scattering medium scenarios via multi-simulated scattering media learning and reconstruct images in real time with light-speed processing. We achieve 80 Hz imaging of moving objects in dynamic scattering media with decorrelation times $<$ 1 ms, demonstrating applicability in large-field-of-view imaging and incoherent illumination scenarios. This work lays the foundation for high-speed, intelligent optical imaging, advancing practical anti-dynamic scattering techniques. 

\end{abstract}

\maketitle
\section*{\label{sec:level1}Introduction}

The ability to perform optical imaging through scattering media is extremely valuable for numerous applications, from astronomical observations through turbulent atmospheres \cite{davies2012adaptive} to microscopic imaging in turbid tissues \cite{ntziachristos2010going, bilsing2024adaptive}, image transmission \cite{bertolotti2022imaging}, and endoscopic diagnostics \cite{wen2023single}.
However, random scattering-induced perturbations disrupt the optical wavefront, leading to the detection of random speckle patterns. This breaks the traditional point-to-point imaging paradigm \cite{bertolotti2022imaging} and poses a significant challenge to achieving high-fidelity image reconstruction. 

Although scattering appears random, within a specific time window—commonly referred to as the speckle correlation time, ${\tau _w}$—the scattering process remains deterministic. Leveraging this property, various techniques have been developed for optical focusing and imaging through scattering media, including transmission matrix (TM) measurement \cite{PhysRevLett2010measure, popoff2010image,boniface2020non}, optical phase conjugation (OPC) \cite{yaqoob2008optical, xu2011time}, iterative wavefront shaping (IWS) \cite{mosk2012controlling,katz2012looking, sciadv2021Tomer,Sun2024optica}, optical memory effect (OME) \cite{bertolotti2012non, katz2014non,zhu2022large}, and artificial neural networks (ANN) \cite{Li2018imaging,Li2018deep, Zhu2021, Mengap2019, Zhang2023, gao2021distortion}. However, real-world scattering media, such as fog, biological tissues, and turbid fluids, are inherently dynamic, leading to rapid temporal decorrelation of optical information.

To address this challenge, TM and IWS methods employ advanced modulation devices \cite{yang2021anti,tzang2019wavefront} and high-efficiency algorithms \cite{valzania2023online,sciadvBrandon2023,He2024pr} to minimize time consumption, ensuring that the scattering medium remains quasi-static within an operational cycle ${\tau _o}$ and thus meeting the one-to-one measurement and compensation requirements, namely, ${\tau _o<\tau _w}$. Consequently, the number of spatial modes that can be controlled before speckle decorrelation occurs is a key parameter determining the performance of wavefront shaping. In contrast, OPC can obtain the correct wavefront without iteration, enabling a higher number of controllable optical modes in the replayed wavefront. This method achieves an average control time per spatial mode several orders of magnitude shorter than the other two approaches \cite{sciadvadd9158}, at the cost of increased system complexity. Currently, OPC combined with ultrasound-guided stars \cite{liu2015optical} is widely used for focusing through scattering media \cite{Liu17optica, Wang15optica}. Like TM and IWS methods, OPC remains limited by speckle decorrelation.

Data-driven deep learning \cite{fu2024adaptive, liu2023deep} has also been employed for end-to-end object image reconstruction, yet these networks exhibit limited generalization capabilities and can typically accommodate only a few dozen mapping relationships \cite{zhang2024gen}, whereas leveraging ballistic components has the potential to extend generalization \cite{liu2024learning}. Adaptive \cite{Luo2021} and self-supervised dynamic learning \cite{li2024self} methods provide partial compensation for speckle decorrelation, proving effective in slowly varying conditions but inadequate for highly dynamic scattering media with rapid fluctuations. In addition, various techniques, including the optical shower-curtain effect \cite{Edrei2016}, bispectrum analysis \cite{Hwang2019bispectrum}, coherent averaging \cite{Hwang2023coherent}, virtual reference wave technology \cite{luosingle}, and fluorescence-modulated signal correlation \cite{ruan2020fluorescence}, have been explored for imaging through non-static media \cite{wang2021nc}. These methods typically require continuous multi-frame measurements, complex experimental setups, and slow reconstruction algorithms, prolonging the imaging process and making them effective only within a narrow range. Furthermore, they often necessitate that the target object remains stationary during the measurement process or must move less than the OME range between two successive frames\cite{jauregui2022tracking}. Thus far, no existing method enables real-time reconstruction of moving objects without being affected by speckle decorrelation.

Leveraging the parallelization, multitasking, low latency information processing, and computational capabilities of light, the ODNNs \cite{sciencea2018} have recently excelled in beam shaping \cite{veli2021terahertz}, privacy encryption \cite{yang2024apn}, communication engineering \cite{luo2019design}, and image reconstruction \cite{luo2022computational,LAM2023010005,li2023opt}. Initially, the Ozcan group \cite{luo2022computational} pioneered ODNN-based imaging through diffusers in the THz regime using 3D-printed media matched to simulations. These studies, rooted in optical computing, focused on proof-of-concept demonstrations without addressing real-world scattering or the physics principles underlying the imaging process. Recently, ODNNs has been explored for imaging through {\color{black} real-world} scattering media \cite{sciadvadn2024zhang, NP2025gumin}. However, these studies have used training datasets obtained from real experimental scenarios, with imaging conducted under static scattering conditions. Meanwhile, the acquisition of experimental datasets constitutes a highly resource-intensive and time-consuming endeavor, imposing constraints on the generalizability and practical deployment of the resulting trained networks, making this approach unsuitable for imaging through dynamic scattering media. Currently, a fundamental yet unresolved challenge remains: can a fixed optical modulation, trained solely on simulated scattering media, enable imaging through real-world dynamic scattering media? Addressing this challenge is essential for advancing real-time imaging methods capable of mitigating dynamic scattering disturbances, thereby enabling the practical application of anti-scattering imaging technologies.

In this work, we propose a scheme for real-time imaging of moving objects through dynamic scattering media using ODNNs. Fundamentally, the method strives to determine a fixed modulation module that counteracts the perturbations introduced by time-varying scattering, providing a novel strategy for addressing dynamic scattering. 
The proposed scheme is all optical, scan-free, guide-star-free, controlled-illumination-free, and post-processing-reconstruction-free. Most importantly, it does not require any experimental datasets for training. 
Instead, ODNNs are trained under fully simulated objects and scattering media, enabling generalization to real scattering environments, where the scattering media are generated by convolving random matrices with Gaussian kernels. 
By employing a joint training strategy with hundreds of simulated scattering media, we obtain ODNNs robust to diverse scattering conditions. The randomness in training enhances generalization to unseen scattering media, coupled with the ODNNs' light-speed processing enabling real-time object reconstruction at the detection plane. 
By mapping the optimized ODNNs onto spatial light modulators, we achieve real-time imaging through dynamic single-layer and double-layer ground glass, as well as multilayer tracing paper, A4 paper, and vinyl electrical tape. As a proof-of-concept demonstration, a motorized translation stage moves target objects, including digits, letters, special symbols, Chinese characters, patterns, and natural images. 
Moreover, our framework enables imaging beyond the OME range and is adaptable to incoherent illumination, with promising implications in turbulence and biomedical imaging.

\section*{\label{section2}Results}

\subsection*{The principle of the scheme}

\begin{figure*}[t]
    \begin{center}    
    \includegraphics[width=0.8\columnwidth]{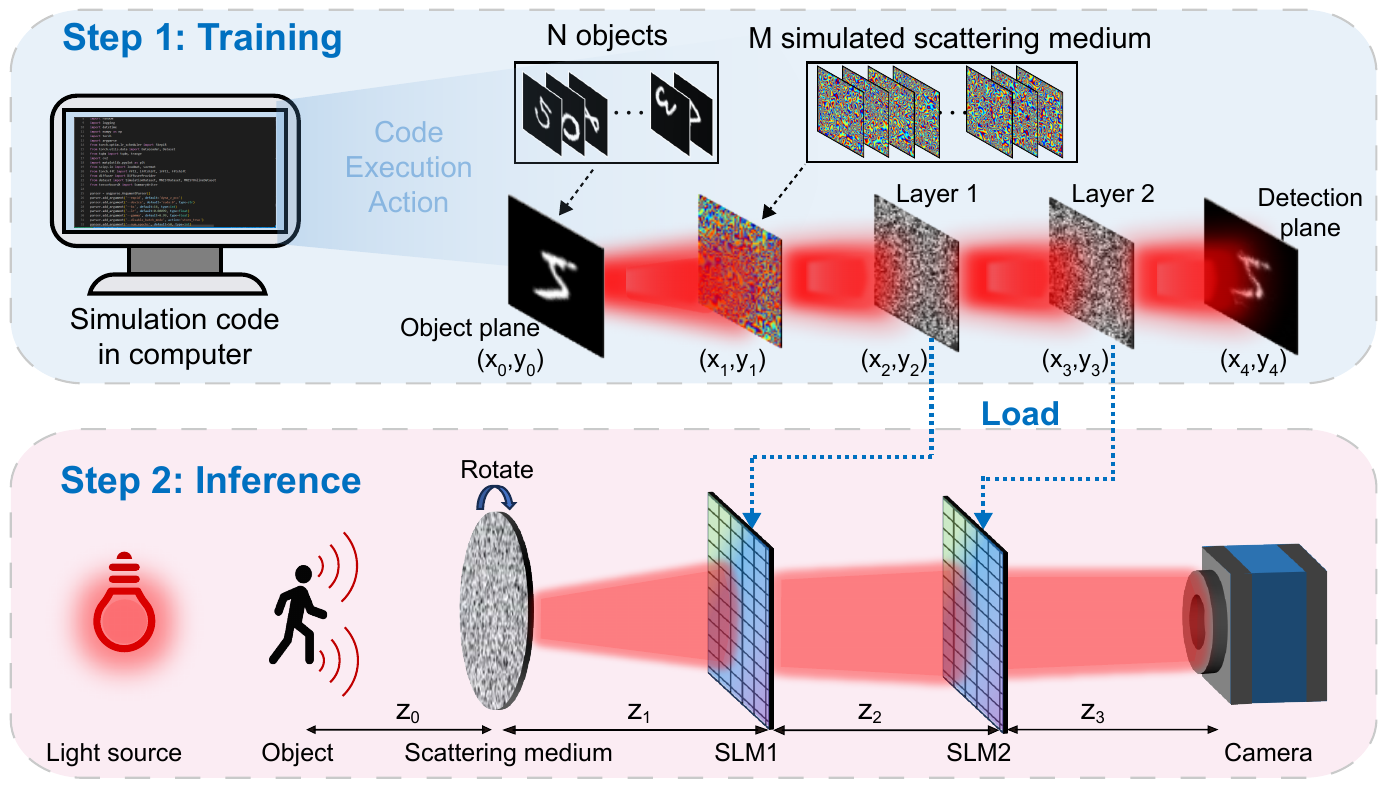}
    \caption{{\textbf{Mechanism for image reconstruction through scattering medium using optical diffraction neural network.} Step 1: The simulation code executes the light field propagation process by loading $N$ different MNIST objects and $M$ simulated scattering media to optimize the ODNNs (illustrated with two layers). 
    Step 2: The trained ODNNs are then loaded onto the phase-type SLMs to achieve real-time imaging through real dynamic scattering media. 
    {\label{fig1-1}}  }}
    \end{center}
\end{figure*}

Fig. \ref{fig1-1} presents the diagram of the proposed method, exemplified with a two-layer ODNN. 
The goal is to image a hidden  object that is located behind a strongly dynamic scattering medium in Step 2. This is achieved through two processes: digital training and optical inference.

We first establish an ODNN-based light propagation model (Step 1). The input-output relationship of the model can be described in the form of a transmission matrix, expressed as 
\begin{equation}
\begin{aligned}
	{E^\mathrm{out}} = T{E^\mathrm{in}}
	\label{eq1}
\end{aligned}
\end{equation}
where $T$ is the transmission matrix of the imaging system, $E^\mathrm{out}$ and $E^\mathrm{in}$ are the output (detection plane) and input  (object plane) light fields, respectively. Notably, the transmission matrix $T$ can be expressed as $T = {T^\mathrm{ODNN}}{T^\mathrm{SM}}$, where $T^\mathrm{SM}$ and $T^\mathrm{ODNN}$ represent the transmission matrices through the scattering medium and the ODNN, respectively.

If $T^\mathrm{SM}$ remains constant, it is always possible to find its pseudo-inverse matrix $(T^\mathrm{SM})^{+}$, allowing $T^\mathrm{ODNN}=(T^\mathrm{SM})^{+}$ to ensure that the transmission matrix $T$ of the model is approximately the identity matrix $I$. This is the basis for the practice of the wavefront shaping methods (TM, IWS, OPC) under static or quasi-static scattering scenarios. Nevertheless, when $T^\mathrm{SM}(t)$ varies with time, such as under different scattering media or dynamic scattering conditions (under conditions that violate the quasi-static assumption), a fundamental condition needs to be satisfied: the ODNN's transmission matrix $T^\mathrm{ODNN}$ shall be constructed to ensure that the output image intensities $I^{out}(t)=\left| {E^\mathrm{out}(t)} \right|^2$ and input image intensities ${\left| {E^\mathrm{in}} \right|^2}$  are approximately equal, such that 
\begin{equation}
\begin{aligned}
    I^{out}(t) &= {\left| {{T^\mathrm{ODNN}}T^\mathrm{SM}(t) E^\mathrm{in}} \right|^2} \\
    &\approx {\left| {E^\mathrm{in}} \right|^2}
	\label{eq2}
\end{aligned}
\end{equation}
holds. Reference \cite{mididoddi2025threading} investigates a related problem by identifying stable transmission channels within multilayer scattering media that exhibit only localized structural changes over the wavefront shaping timescale. Here, our work focuses on scenarios—whether single- or multilayer—in which the entire scattering region undergoes significant and global changes within the same timescale. Consequently, no stable transmission channels exist in any of the cases we consider. Under such conditions, there has been no physics-based theoretical report on the existence of $T^\mathrm{ODNN}$ satisfying Eq. (\ref{eq2}), nor on the conditions under which it exists.

In this case, the ODNNs, physics-informed neural networks trained via supervised learning, provide an effective solution to the problem of dynamic scattering media. 
In our work, the expected output image at the detection plane corresponds to the input image at the object plane. That is, the input object pattern serves as the label for supervised learning (Figs, S14(b) and S33 present examples in which transformed object images serve as training labels). 
In simulations, the propagation of the light field between adjacent planes is modeled using the Fresnel diffraction \cite{goodman2005introduction}, while the simulated scattering medium is generated by convolving a random matrix with Gaussian kernels of varying radii {\color{black}$\sigma_K$}. 
The objective of {\color{black} training} is to optimize a set of ODNNs, i.e., a set of phase patterns $\{\phi_\mathrm{1}(x_2,y_2), \phi_\mathrm{2}(x_3,y_3)\}$ distributed on the interval $(0, 2\pi]$ for modulating the phase of the light field, by using $N$ objects and $M$ simulated scattering media as inputs, so as to generate $T^\mathrm{ODNN}$ that satisfies Eq. (\ref{eq2}). 
As a result, the following equation 

\begin{equation}
\begin{aligned}
 &~~~~ \mathop{\operatorname*{arg\,min}}\limits_{\phi_1, \phi_2} 
   -\frac{1}{M N} \sum_{i=1}^{M} \sum_{j=1}^{N} \mathcal{P} \left( 
      \left| E_j^{\mathrm{in}} \right|^2,\  
      \left| E_{ij}^{\mathrm{out}} \right|^2 
    \right) \\
  & = -\frac{1}{M N} \sum_{i=1}^{M} \sum_{j=1}^{N} \mathcal{P} \biggl(
      \left| E_j^{\mathrm{in}} \right|^2,\ 
      \left| T^{\mathrm{ODNN}}(\phi_1, \phi_2) T_{ij}^{\mathrm{SM}} E_j^{\mathrm{in}} \right|^2
    \biggr)
\end{aligned}
\label{eq3}
\end{equation}
is optimized, where $i = 1,2, \cdots ,M$ denotes a set of time-discrete scattering media, $j = 1,2, \cdots ,N$ denote the different input image fields used for training. $\mathcal{P}(\cdot)$ denotes the Pearson correlation coefficient (PCC) \cite{Benesty2009}, {\color{black} as defined in Eq. (\ref{eq4}).}  
The well-trained ODNNs can reconstruct input object images even in the presence of unknown objects and scattering media. This shows that the ODNN has established a relationship that approximately fulfills Eq. (\ref{eq2}). 
Subsequently, mapping phase patterns $\phi_\mathrm{1}$ and $\phi_\mathrm{2}$ onto the SLMs in Step 2 enables optical inference.

\subsection*{The numerical results of the scheme}

\begin{figure*}[t]
    \begin{center}
    \includegraphics[width=0.8\columnwidth]{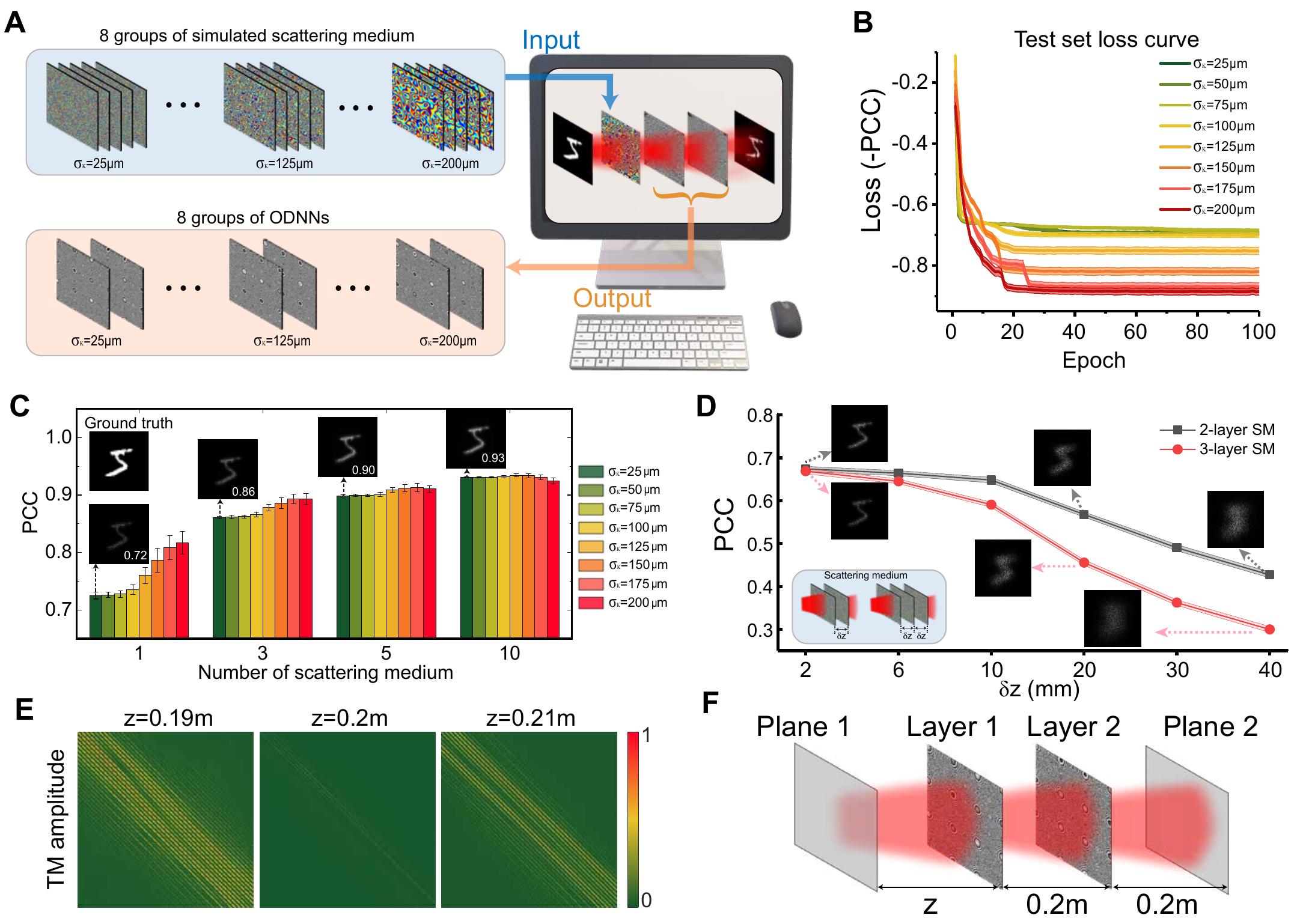}
    \caption{\textbf{Simulation results are obtained under simulated scattering media with different scattering intensities.} \textbf{A} Eight ODNNs are trained using eight sets of simulated scattering media datasets, each generated by convolving with different kernel sizes. \textbf{B} The loss function curves of the test set are obtained during the training of ODNNs under eight different simulated scattering media respectively. \textbf{C} The ODNNs trained on the $\sigma_K$= 25 $\mu$m scattering media are tested on eight types of simulated scattering media. From left to right, the images represent the quality of the reconstructed images obtained by summing the intensities of results recovered under 1, 3, 5, and 10 different scattering media. \textbf{D} Imaging results under two-layer and three-layer simulated scattering media ($\sigma_K$= 25 $\mu$m).  The ODNNs are trained with a single-layer simulated scattering medium at $\sigma_K$= 25 $\mu$m. SM: scattering media. \textbf{E} Amplitude patterns of the transmission matrix from plane 1 to plane 2 at different z, where the measurement setup is shown in \textbf{F}. The ODNN is trained on the scattering media with $\sigma_K$= 25 $\mu$m.}
    {\label{fig1-2}}  
    \end{center}
\end{figure*}

Based on the aforementioned principles, we generated eight sets of simulated scattering media training datasets by convolving random matrices with Gaussian kernels of eight different sizes $(\sigma_K = 25,50, \cdots ,200~\text{\textmu m})$. Each dataset contains $M=1,000$ randomly generated scattering media, with the convolution kernel sizes corresponding to 1,2, $\cdots$, and 8 simulated pixel units (see Supplementary Note 3 for details). Using these datasets, we trained eight corresponding sets of ODNNs, as illustrated in Fig. \ref{fig1-2}A. The object training set is derived from the MNIST \cite{lecun1998gradient} dataset ($N = 20,000$). For evaluation, we employed a test set consisting of 5,000 unseen objects and 50 unseen scattering media, sampled from the same distribution as the training set. The test set loss function curves over training epochs are shown in Fig. \ref{fig1-2}B. As convolution kernels $\sigma_K$ increases, that is, the scattering intensity gradually decreases, the quality of the final converged results increases, consistent with expectations.

To evaluate the generalization of ODNNs, we use the ODNNs trained under $\sigma_K$= 25 $\mu$m simulated scattering media to reconstruct images under different scattering intensities. The fidelity of the reconstructed images, measured by the Pearson correlation coefficient, is shown in the leftmost column of Fig. \ref{fig1-2}C. Notably, fidelity is higher under weak scattering conditions (large $\sigma_K$), and the best performance does not occur at ${\color{black}\sigma_K}$= 25 $\mu$m, consistent with physical principles rather than training set constraints. To demonstrate reconstruction under dynamic scattering conditions, we sum the intensity of multiple reconstructed images from multiple static scattering scenarios, simulating extended exposure time or employing scattering media with faster decorrelation times.
By summing the reconstructed images obtained under five different scattering media (under the same statistical distribution), the overall image fidelity exceeds 0.9 (more results in Fig. S27). This indicates that dynamic scattering media, often considered an obstacle, can instead be leveraged to enhance imaging quality. The reason lies in the ODNN’s ultrafast information processing, which allows real-time acquisition of corrected images at the detection plane. Temporal integration over a certain period functions as statistical averaging, effectively reducing noise and improving fidelity. This further suggests that our method remains unaffected by scattering medium decorrelation.

Fig.\ref{fig1-2}C also demonstrates that ODNNs can successfully reconstruct object images under a single-layer scattering medium, regardless of the scattering strength. To further investigate its performance limits, we {\color{black}evaluated the reconstruction capability of ODNNs} under two-layer and three-layer scattering media, as shown in Fig. \ref{fig1-2}D. 
As observed, the image quality degrades with increasing layer number and spacing, as expected. This degradation may also stem from discrepancies between the training and testing conditions.
We refined the model during training; however, even with additional ODNN layers, it failed to optimize an ODNN that satisfies Eq. (\ref{eq2}), as shown in Fig. S28.
This suggests that the limitation arises from fundamental physical and mathematical constraints rather than model imperfections or network capacity limitations.

This finding reveals a fundamental trade-off between ODNNs generalization and scattering intensity. Essentially, a single-layer simulated scattering medium functions as a phase screen without introducing amplitude perturbations. 
As the number of scattering layers and their spacing increase, the effect of phase perturbations gradually spreads, forming a random speckle pattern that undergoes further modulation in the subsequent layer, resulting in an increasingly complex scattering outcome. 
This transition shifts the scattering model from simple surface scattering to a more complex volumetric scattering, pushing the boundary of the solution regime described by Eq. (\ref{eq2}). In such scenarios, a reliable solution would be to measure the transmission matrix, enabling one-to-one compensation. This is also the current method for imaging through complex scattering media such as multimode fibers \cite{wen2023single, kupianskyi2024all, butaite2022build}.


To elucidate the imaging mechanism of ODNNs through scattering media, we measured the optical field transmission matrix from plane 1 to plane 2 following the schematic in Fig. \ref{fig1-2}F. Fig. \ref{fig1-2}E shows the transmission matrix amplitude patterns at different propagation distances z (from plane 1 to layer 1). Remarkably, when z=0.2m, the transmission matrix approaches a diagonal form. This distance coincides precisely with the separation between diffraction layer 1 and the scattering medium in the simulation, indicating that the ODNN focuses on the scattering medium plane. This behavior is consistent with the principle of imaging based on the optical shower-curtain effect \cite{Edrei2016}. Specifically, the ODNN targets the rear surface of the scattering medium and exploits the spatial correlation between its front and rear sides to reconstruct the image. When the scattering strength is relatively weak, this front–rear correlation remains high, thereby preserving imaging fidelity. However, as the scattering strength increases, the correlation progressively diminishes, ultimately leading to the breakdown of the imaging scheme (see Supplementary Note 9.2 for further details).

\subsection*{Imaging dynamic objects through dynamic scattering media}

\begin{figure*}[htbp]
    \begin{center}
    \includegraphics[width=0.8\columnwidth]{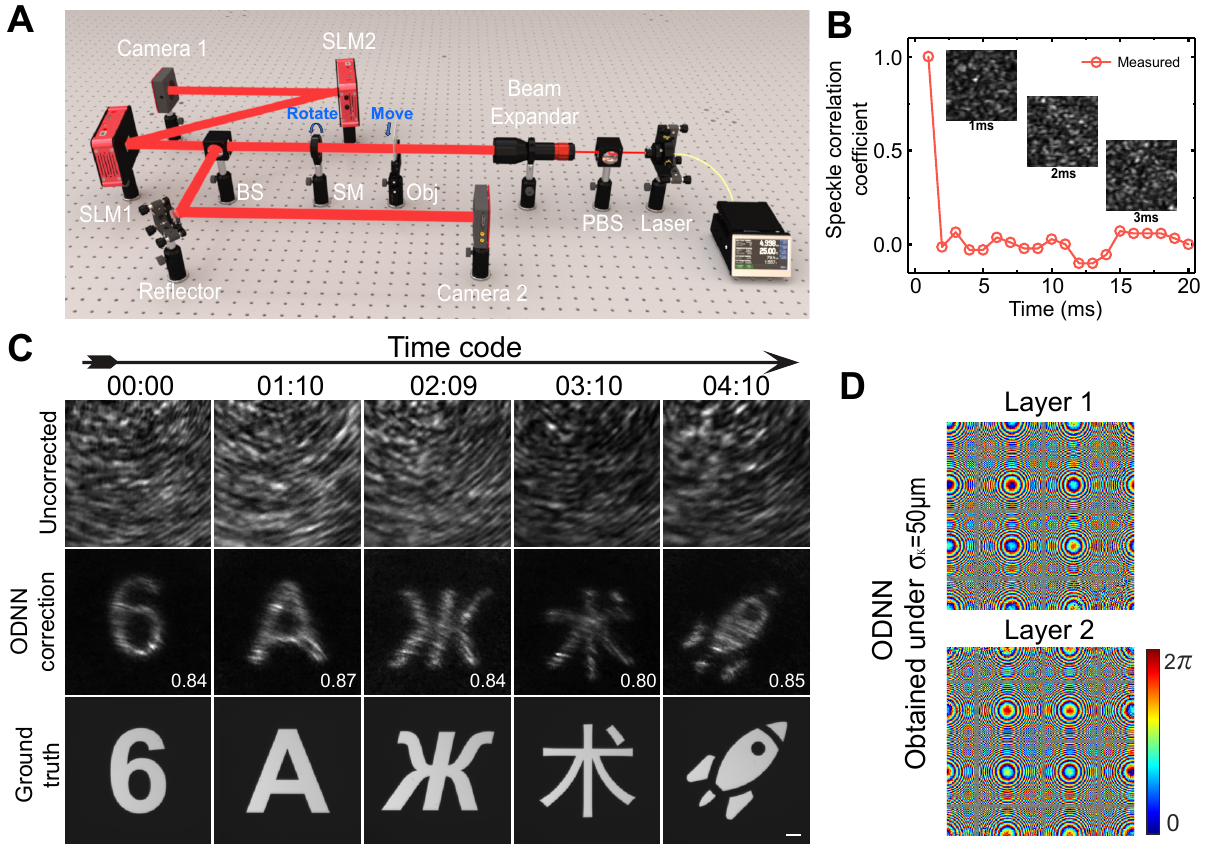}
    \caption{{\textbf{Experimental results of imaging dynamic objects through dynamic scattering media.} 
    \textbf{A} Experimental setup diagram in coherent illumination scenarios. PBS: polarizing beamsplitter, Obj: object, SM: scattering medium, BS: beam splitter, SLM: spatial light modulator. 
    \textbf{B} The decorrelation curve of the dynamic scattering medium. 
    \textbf{C} The imaging results under the dynamic (rotating) 120-grit diffuser. The first row shows the results detected by camera 2 without ODNNs correction. The second row displays the results detected by camera 1 after ODNNs correction. The third row shows the ground truth. Time code, seconds : frames. 30 frames per second. See the video in movie S1. 
    \textbf{D} The diffraction layers loaded onto the spatial light modulators, which are obtained by training under the simulated scattering medium with $\sigma_K=\mathrm{50~\mu m}$.  
    Scale bars, 0.5 mm. }}
    {\label{fig2}}  
    \end{center}
\end{figure*}

As an experimental proof of the concept, we established the experimental setup depicted in Fig. \ref{fig2}A to perform imaging of five types of objects through a highly scattering diffuser (120-grit, Thorlabs). The ODNNs (Fig. \ref{fig2}D) used here are trained on simulated scattering media with $\sigma_K=\mathrm{50~\mu m}$. The scattering medium is controlled by a custom-built rotating stage (Fig. S8) with a decorrelation time of less than 1 ms, as shown in Fig. \ref{fig2}B. The objects are translated using a linear stage at an average speed of 5 mm/s, with a fixed camera exposure time of 33.3 ms. The reconstructed images are presented in Fig. \ref{fig2}C, demonstrating that ODNNs effectively modulate the light field, allowing camera 1 to capture real-time reconstructed object images (second row). In contrast, images acquired by camera 2 along the same propagation {\color{black}distance} appear as random speckle patterns (first row). 
These uncorrected patterns display noticeable rotational motion blur instead of the high-contrast speckle patterns, indicating significant changes in the scattering medium within a single exposure time. The ODNNs-corrected results not only clearly reveal the object’s image but also capture positional and motion information, as shown in movie S1. Compared to imaging results under static scattering media (see Fig. S35), in dynamic scenarios, each single-frame exposure on the camera accumulates multiple independent and uncorrelated reconstruction results, effectively suppressing noise outside the object region and enhancing image quality.

\begin{figure*}[t]
    \begin{center}
    \includegraphics[width=0.80\columnwidth]{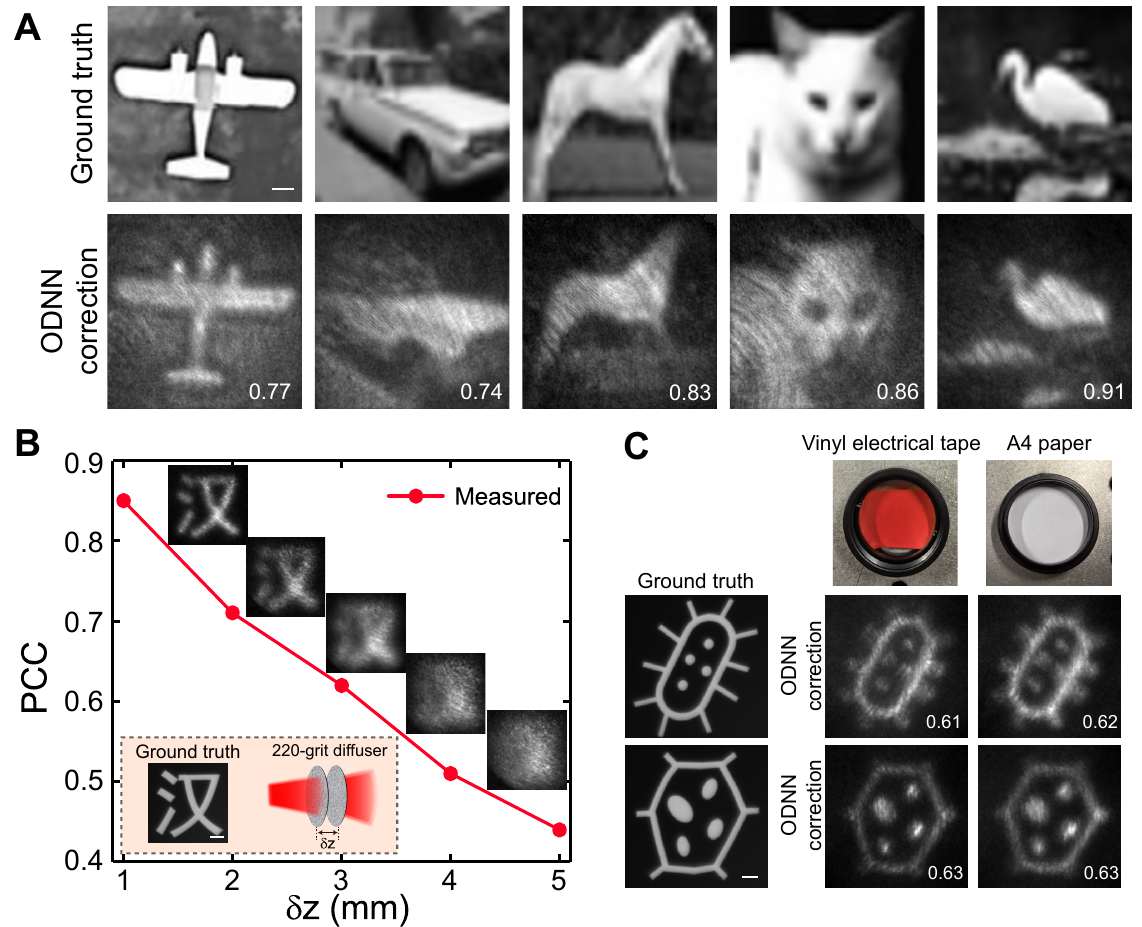}
    \caption{{\textbf{Experimental results under natural images and common scattering media.} 
    \textbf{A} Experimental results under natural images, where the images are projected by a DMD. Scattering medium: 120-grit diffuser. 
    \textbf{B} Imaging results through double-layer 220-grid diffuser with different inter-layer spacings. 
    \textbf{C} Imaging results through dynamic A4 paper (80g/$\mathrm{m^2}$) and vinyl electrical tape(3M company).  
    Scale bars, 0.5 mm. }}
    {\label{fig3}}  
    \end{center}
\end{figure*}

To demonstrate the generalizability of ODNNs to different types of images, we conducted imaging experiments using natural images from the OPT-Aircraft \cite{aircraft2020} and CIFAR-10 \cite{Krizhevsky2009LearningML} datasets.  The results are presented in Fig. \ref{fig3}A, where these patterns are displayed using a digital micromirror device. Fig. \ref{fig3}C shows imaging results conducted using common opaque scattering media (standard A4 paper and vinyl electrical tape) as scattering media. 
Fig. S40 shows the reconstruction results of eight ODNNs under 120-grit ground glass. 
The results demonstrate that object images are successfully retrieved in all cases. The variations in reconstruction quality stem from intensity fluctuations induced by dynamic scattering media (see Supplementary Note 8.3 for details). 
Notably, despite being trained on MNIST handwritten digits, ODNNs exhibit remarkable imaging capabilities for non-digit objects, distinguishing our method from conventional deep learning methods. This suggests that ODNNs perform physics-based correction and compensation rather than merely memorizing patterns.

Furthermore, we {\color{black}stacked} two 220-grit diffusers and varied the inter-layer spacing to create a scattering medium with tunable thickness and experimentally explored the imaging limits of ODNNs. 
We found that when the layer spacing is $\delta_z=3~\mathrm{mm}$, the reconstructed image became blurred (Fig. \ref{fig3}B). However, in simulations, significant degradation only occurred when  $\delta_z$=3--4 cm (Fig. \ref{fig1-2}D). 
This discrepancy arises from the limitations of the simulated scattering model, as phase screens represent an abstract, simplified approximation that cannot fully capture the complexity of real scattering processes. As a result, physical thickness (or range) becomes the dominant constraint on imaging performance in simulations (see different $Z_1$ in Fig. S10). 
In contrast, in experimental settings, the primary limitation stems from scattering strength--well before thickness becomes a critical factor.
Fig. S21(a) presents a more comprehensive results using tracing paper as an example, demonstrating that image quality degrades with increasing numbers of layers (scattering strength) and inter-layer spacing. For two layers of tracing paper, the spacing is also limited to within a few millimeters.

To quantitatively assess the upper limit of our method’s robustness against scattering, we systematically stack increasing layers of tracing paper, A4 paper, and vinyl electrical tape to create media of varying scattering strength. For consistent comparison, scattering strength is quantified using optical depth, defined as the ratio of physical thickness to the transport mean free path. As shown in Figs. S21(c) and (d), the images become unrecognizable at optical depths of 1 (tracing paper), 1.54 (A4 paper), and 1.19 (vinyl electrical tape), respectively, indicating that our method remains effective within approximately 1--2 transport mean free paths.

Additional experimental results and detailed analyses are available in the Supplementary Materials. These include standard optical measurements (Supplementary Note 7), a robustness analysis (Supplementary Note 8), as well as in-depth investigations of the imaging mechanism, boundary, ODNN morphology and spatial resolution limits (Supplementary Notes 9–11).

\begin{figure*}[htbp]
    \begin{center}

    \includegraphics[width=0.8\columnwidth]{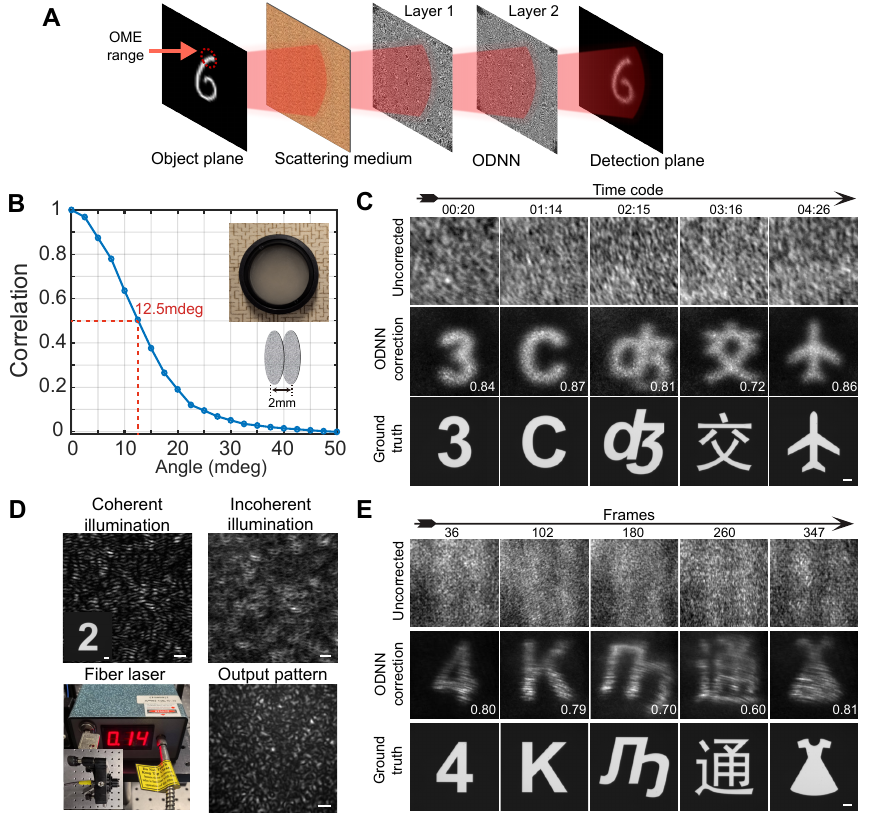}
    \caption{{\textbf{Experimental results under large field-of-view and incoherent illumination conditions.} 
    \textbf{A} Schematic diagram of large field-of-view imaging . 
    \textbf{B} The decorrelation curve of the double-layer 220-grit diffuser, and the inset shows the visual representation of the scattering medium. 
    \textbf{C} Real-time large field-of-view imaging of moving objects through dynamic scattering media. Time code, seconds : frames. {\color{black}30 frames per second.} See video in movie S2.   
    \textbf{D} First row: The speckle patterns generated by the digit '2' passing through the static scattering medium under coherent and incoherent illumination.  Second row: Photograph of the fiber laser (left) and the output intensity pattern under static conditions (right). 
    \textbf{E} Imaging of objects under incoherent illumination from a multimode fiber through a dynamic 120-grit diffuser.
    See video in movie S4. Scale bars, 0.5 mm. }}
    {\label{fig4}} 
    \end{center}
\end{figure*}

\subsection*{Large-field-of-view imaging of dynamic objects through dynamic scattering media}

To demonstrate large-field-of-view (FOV) imaging capability in the dynamic scattering medium scenario, we stack two pieces of 220-grit ground glass, separated by a 2-mm-thick retaining ring, to form a thicker scattering medium with higher scattering intensity.
The OME range of the double-layer 220-grit scattering medium is measured by tilting the incident laser angle and recording the speckle decorrelation \cite{Sun2024optica, katz2014non}, with the decorrelation curve shown in Fig. \ref{fig4}B, indicating that half of the OME range is approximately 12.5 mdeg.
When object distance $Z_0$ = 1 cm, the diameter of FOV within a single OME range is around $4.36~\mathrm{\mu m}$.
Fig. \ref{fig4}C shows the imaging results of five different objects through dynamic scattering media at $Z_0$ = 1 cm (with results under static scattering shown in Fig. S37, see video in movie S2). From left to right, the object heights are 2.71 mm, 2.75 mm, 3.43 mm, 3.09 mm, and 3.21 mm, respectively.
Correspondingly, the imaging FOV is expanded by factors of 622, 631, 787, 709, and 736 (see Supplementary Note 13.2 for details). As $Z_0$ further decreases, imaging remains achievable, and the magnification of the FOV will further increase.

\subsection*{Generalize to incoherent illumination}

Incoherent light sources are more widely used due to their affordability and accessibility, and are particularly suited for biomedical imaging applications, such as white light and fluorescence.
Incoherent imaging systems are linear with respect to intensity rather than complex amplitude \cite{goodman2005introduction}. Consequently, the scattering pattern changes from a random speckle pattern to a low-contrast pattern that contains the object's structural information (the first row in Fig. \ref{fig4}D). Although light from a spatially incoherent extended object does not have a well-defined wavefront, the light from each point on the object does have a defined wavefront, which can be modulated and compensated by the well-trained ODNN with coherent illumination. This feasibility reveals that the learned ODNN can generalize well to real-time compensation of unknown distorted illumination wavefronts.
It is important to note that uneven illumination intensity may degrade the results at each moment, but temporally accumulated illumination will become more uniform, allowing the detection results to converge to those obtained under coherent illumination.

In our experiment, two different incoherent light sources are used to illuminate the objects: one is directly emitted from a multimode optical fiber port (the second row in Fig. \ref{fig4}D) and the other is generated by modulating a rotating ground glass (Figs. S38 and S39, movie S3). Fig. \ref{fig4}E illustrates the reconstruction results obtained under incoherent illumination provided by a rapidly jittering multimode fiber. The camera exposure time is set to 12.5 ms, corresponding to an acquisition rate of 80 frames per second. See video in movie S4.

\section*{\label{section3}Discussion }

We demonstrate real-time imaging of moving objects through real dynamic scattering media using ODNNs trained on simulated scattering media.  The method does not require either the object or the scattering medium to remain quasi-static in the imaging process. We achieve an imaging rate of 80 Hz in dynamic scattering scenarios with decorrelation times of less than 1 ms. With sufficient illumination and more advanced cameras, the frame rate can be further improved. Our method is applicable to incoherent illumination scenarios and enables imaging independent of the optical memory effect. This provides a novel approach for imaging fluorescent objects embedded within scattering media.  
Mathematically, the feasibility of our method depends on the existence of $T^\mathrm{ODNN}$ that satisfies Eq. (\ref{eq2}), while physically, it is attributed to the optical shower-curtain effect. 
When encountering multi-layer or volumetric scattering, the complexity of $T^\mathrm{SM}(t)$  \cite{mididoddi2025threading} in Eq. (\ref{eq2}) imposes stringent constraints on identifying a valid $T^\mathrm{ODNN}$, potentially rendering it nonexistent. The results imply that more advanced methods may be required to resolve the end-to-end real-time imaging in dynamic volumetric scattering media. Recently, nonlinear diffractive neural networks have demonstrated potential advantages in tasks such as image classification, recognition, and optical encoding \cite{wang2023image}. We anticipate that this nonlinear representational capacity may enhance imaging quality in our context. 
More significantly, this enhanced capacity could potentially help push the fundamental imaging boundaries of our method, suggesting a promising avenue for future exploration.


In simulations, the imaging performance begins to converge when the training set contains around 6,000 objects (with 1,000 simulated scattering media), and further increases do not significantly improve quality (Fig. S5(a)). While this number may be influenced by factors such as batch size and learning rate, it indicates that the overall required sample size is relatively small. Additionally, a few hundred simulated scattering media (with 20,000 objects) are sufficient to form a training set (Fig. S5(b)), as they provide enough randomness. 
In our model, a single-layer ODNN is inadequate  because it cannot simultaneously resist interference from various scattering media and perform equal-sized and upright imaging. This limitation is evident from the abrupt drop in performance illustrated in Fig. S5(d). In contrast, a two-layer ODNN can achieve good performance.  Adding more than two layers can  enhance image quality. Note that strongly scattering media may lead to the failure of the training process of the multi-layer ODNNs. One feasible approach is to optimize a subset of the diffraction layers at a time. Fig. S5(e) demonstrates that when each neuron is connected to $75\%$ of the neurons in the subsequent layer (determined by the propagation distance), the imaging quality reaches a plateau and does not improve significantly. Conversely, if the connectivity falls below this threshold, the PCC decreases noticeably. Therefore, our results reveal that enough optical neuron connections are necessary to enhance the expressiveness of the ODNNs to achieve better imaging quality.


Experimentally, the proposed scheme employs a commercially available liquid crystal-based spatial light modulator to function as a ODNN. Due to the relatively large size of liquid crystal molecules (tens of micrometers), we increase the interlayer spacing to ensure sufficient neuron connectivity between layers. The fabrication and application of subwavelength-structured metasurfaces can significantly enhance neuron density while reducing the overall system volume. 
In our scheme, the trained ODNNs act as a fixed modulation component, eliminating the need for programmable control, which substantially simplifies the fabrication process. 
Fig. S11 demonstrates the system’s robustness against variations of propagation distance. 
Fig. S17 shows that the ODNNs with only 2-bit depth achieve satisfactory imaging performance, highlighting their great potential for optical system integration. 
Furthermore, additional degrees of freedom, such as wavelength, polarization, and orbital angular momentum, can be incorporated to further improve imaging quality, enhance generalization capabilities, and expand potential application scenarios. While orbital angular momentum is a subset of the spatial basis manipulated by ODNN, we suggest that explicitly introducing OAM modes--e.g., through spiral phase structures-could offer additional structural control. This may enable enhanced functionalities such as edge enhancement \cite{sciadvadn2024zhang} or increased robustness \cite{Meglinski2024, khanom2024twists} to scattering, which could be beneficial in certain imaging tasks. 
Although our current experimental demonstration is conducted in a transmission setup, we emphasize that our method is fully compatible with reflective setups. In the scenario of incoherent illumination, the object is placed between two dynamic scattering media—a configuration that can be equivalently implemented in a reflective setup.


In a broader context, a factor contributing to the feasibility of this method is the ability of the simulated scattering medium to effectively replicate the scattering behavior of real scattering media, such as ground glass, A4 paper, etc. Although it relies on a basic phase screen model that is insensitive to polarization and beam incident direction, it can still partially reflect the effects of scattering on the optical field \cite{goodman2007speckle}. The results suggest that although the scattering process is complex and requires high-dimensional parameters for full characterization, scattering speckles can be viewed as projections from this high-dimensional space.
By continuously optimizing the optical diffraction neural network, it can effectively learn strategies to mitigate scattering effects.  
Our approach strikes a favorable balance between data-driven and physics-driven methods \cite{karniadakis2021physics}, ensuring robust out-of-distribution generalization and efficient resistance to dynamic scattering.

\section*{\label{section3}Methods }

\subsection*{Experimental setup}

The complete experimental setups are detailed in Fig. S1. In these experiments, the objects are placed 10 cm behind the scattering medium.
The coherent illumination source is generated by a Bragg-reflection single-frequency laser ($\lambda=780~\mathrm{nm}$). The incoherent illumination sources include two types: one produced by modulating the Bragg-reflection single-frequency laser with a rotating ground glass and the other emitted from the output port of a multimode optical fiber ($\lambda=785~\mathrm{nm}$).
The distances between the scattering medium, SLM1, SLM2, and the camera 1 are all set to 20 cm.
The moving objects are achieved using a linear translation stage operating at 5 mm/s, while the dynamic scattering medium is produced by rotational stages (not depicted in the figure). 
{\color{black}In the experiments, movies S1, S2, and S3 are recorded using acquisition software at a frame rate of 30 fps.} 
In Fig. \ref{fig2}C, as well as in movies S1, the camera exposure time is set to 33.3 ms, {\color{black}aligning with the recording frame rate.} 
In Fig. \ref{fig4}C and movies S2 and S3, the camera exposure time is 100 ms, {\color{black}while the recording frame rate remained at 30 fps.} 
In Fig. \ref{fig4}E and movies S4 the camera exposure time is extended to 12.5 ms. Due to the limitations of light field energy and exposure time, the histograms of the results in Fig. \ref{fig4} primarily exhibit values in the lower range. We enhanced these histograms using the \emph{imagesc} function in MATLAB. The results presented in the movies are raw and have not been enhanced.

\subsection*{Simulation implementation}

In the simulation, the illumination wavelength is 780 nm, and all planes (objects, scattering media, diffraction layer1, diffraction layer2, and detection) are $256 \times 256$ pixels, and each pixel is 25 $\mu$m $\times$  25 $\mu$m . The spacing between adjacent layers is set to 10 cm, 20 cm, 20 cm, and 20 cm, respectively. Adjacent optical planes are interconnected via free-space light propagation in air, which is described using the Rayleigh–Sommerfeld equation (see Supplementary Note 2 for details). 

The simulated scattering medium is essentially a set of random phase screens. They are generated by convolving a set of random matrices with Gaussian kernels of different radii $\sigma_K$. These random matrices are generated under a fixed set of parameters (see Supplementary Note 3 for details). 
We generated a total of eight different distributions of scattering media, with the radii of convolution kernels  $\sigma_K$ = 25$\mu$m (1pixel), $\sigma_K$ = 50$\mu$m (2pixels), $\sigma_K$ = 75$\mu$m (3pixels), $\sigma_K$ = 100$\mu$m (4pixels), $\sigma_K$ = 125$\mu$m (5pixels), $\sigma_K$ = 150$\mu$m (6pixels), $\sigma_K$ = 175$\mu$m (7pixels), and $\sigma_K$ = 200$\mu$m (8pixels) respectively. 
The patterns of the simulated scattering medium are shown in Fig. S2. 

Before training, each handwritten digit (object) from the MNIST dataset is first upscaled from $28\times 28$ pixels to $166\times 166$ pixels using bilinear interpolation, then padded with zeros to create images with $256\times 256$ pixels. Random translations within the range of [-60, 60] pixels are applied separately in the x and y directions, along with random rotations between [0, 360) degrees around the center. During training, each input object in a batch are numerically duplicated m times and individually disturbed by a set of m randomly selected scattering media. These distorted fields are then separately forward propagated through the diffractive network. At the output plane, we calculate the Pearson correlation coefficient between the output patterns and labels and use these values to compute the loss function, where Pearson correlation coefficient is defined as: 
\begin{equation}
    P(o,r) = \frac{{\sum {\left( {o - \bar o} \right) \cdot (r - \bar r)} }}{{\sqrt {\sum {{{(o - \bar o)}^2} \cdot \sum {{{(r - \bar r)}^2}} } } }}.
    \label{eq4}
\end{equation}	

where $r$ is the output image of the diffractive network and $o$ is the ground truth, $\bar r $ and $\bar o $ respectively represent their means. The phases $\phi_\mathrm{1}(x_2,y_2)$ and $\phi_\mathrm{2}(x_3,y_3)$ of the ODNNs are constrained within the range of $0$ to $2\pi$ by applying the modulus operation with $2\pi$. During the quantization of ODNNs, for the phase patterns with a maximum phase of 2$\pi$, the range $[0,2\pi]$ was equally divided into $2^{bit~depth}$ steps. For example, the 1-bit {\color{black}quantization} is binary quantization with 0 and $\pi$, while the 2-bit quantized case has $\{0,\frac{\pi}{2},\pi,\frac{3\pi}{2}\}$, i.e., four steps.

Our models are implemented on the PyTorch (v2.0.1+cu117) platform using Python (v3.8.18) with a GeForce RTX 3090 GPU, an Intel Xeon Platinum 8260 CPU, and 128 GB of RAM, running on a Linux operating system. The calculated loss values are backpropagated to update the phases of the diffractive neurons using the Adam optimizer, with a decaying learning rate defined as $Lr=0.99^{epoch}\times0.004$, where \textit{epoch} denotes the current epoch number. With a batch size of $\mathcal{B}=16$, training a typical diffractive network model takes approximately 2.7 hours to complete over 100 epochs (with $m = 10$  scattering media per epoch and  $M = 1,000$ ), using  $N = 20,000$ MNIST objects. This corresponds to a total of $M\times N$ object-speckle pattern pairs used for training.


\bibliography{apssamp}

\begin{thebibliography}{67}%
\makeatletter
\providecommand \@ifxundefined [1]{%
 \@ifx{#1\undefined}
}%
\providecommand \@ifnum [1]{%
 \ifnum #1\expandafter \@firstoftwo
 \else \expandafter \@secondoftwo
 \fi
}%
\providecommand \@ifx [1]{%
 \ifx #1\expandafter \@firstoftwo
 \else \expandafter \@secondoftwo
 \fi
}%
\providecommand \natexlab [1]{#1}%
\providecommand \enquote  [1]{``#1''}%
\providecommand \bibnamefont  [1]{#1}%
\providecommand \bibfnamefont [1]{#1}%
\providecommand \citenamefont [1]{#1}%
\providecommand \href@noop [0]{\@secondoftwo}%
\providecommand \href [0]{\begingroup \@sanitize@url \@href}%
\providecommand \@href[1]{\@@startlink{#1}\@@href}%
\providecommand \@@href[1]{\endgroup#1\@@endlink}%
\providecommand \@sanitize@url [0]{\catcode `\\12\catcode `\$12\catcode
  `\&12\catcode `\#12\catcode `\^12\catcode `\_12\catcode `\%12\relax}%
\providecommand \@@startlink[1]{}%
\providecommand \@@endlink[0]{}%
\providecommand \url  [0]{\begingroup\@sanitize@url \@url }%
\providecommand \@url [1]{\endgroup\@href {#1}{\urlprefix }}%
\providecommand \urlprefix  [0]{URL }%
\providecommand \Eprint [0]{\href }%
\providecommand \doibase [0]{https://doi.org/}%
\providecommand \selectlanguage [0]{\@gobble}%
\providecommand \bibinfo  [0]{\@secondoftwo}%
\providecommand \bibfield  [0]{\@secondoftwo}%
\providecommand \translation [1]{[#1]}%
\providecommand \BibitemOpen [0]{}%
\providecommand \bibitemStop [0]{}%
\providecommand \bibitemNoStop [0]{.\EOS\space}%
\providecommand \EOS [0]{\spacefactor3000\relax}%
\providecommand \BibitemShut  [1]{\csname bibitem#1\endcsname}%
\let\auto@bib@innerbib\@empty
\bibitem [{\citenamefont {Davies}\ and\ \citenamefont
  {Kasper}(2012)}]{davies2012adaptive}%
  \BibitemOpen
  \bibfield  {author} {\bibinfo {author} {\bibfnamefont {R.}~\bibnamefont
  {Davies}}\ and\ \bibinfo {author} {\bibfnamefont {M.}~\bibnamefont
  {Kasper}},\ }\bibfield  {title} {\bibinfo {title} {Adaptive optics for
  astronomy},\ }\href@noop {} {\bibfield  {journal} {\bibinfo  {journal}
  {Annual Review of Astronomy and Astrophysics}\ }\textbf {\bibinfo {volume}
  {50}},\ \bibinfo {pages} {305} (\bibinfo {year} {2012})}\BibitemShut
  {NoStop}%
\bibitem [{\citenamefont {Ntziachristos}(2010)}]{ntziachristos2010going}%
  \BibitemOpen
  \bibfield  {author} {\bibinfo {author} {\bibfnamefont {V.}~\bibnamefont
  {Ntziachristos}},\ }\bibfield  {title} {\bibinfo {title} {Going deeper than
  microscopy: the optical imaging frontier in biology},\ }\href@noop {}
  {\bibfield  {journal} {\bibinfo  {journal} {Nature methods}\ }\textbf
  {\bibinfo {volume} {7}},\ \bibinfo {pages} {603} (\bibinfo {year}
  {2010})}\BibitemShut {NoStop}%
\bibitem [{\citenamefont {Bilsing}\ \emph {et~al.}(2024)\citenamefont
  {Bilsing}, \citenamefont {N{\"u}tzenadel}, \citenamefont {Burgmann},
  \citenamefont {Czarske},\ and\ \citenamefont
  {B{\"u}ttner}}]{bilsing2024adaptive}%
  \BibitemOpen
  \bibfield  {author} {\bibinfo {author} {\bibfnamefont {C.}~\bibnamefont
  {Bilsing}}, \bibinfo {author} {\bibfnamefont {E.}~\bibnamefont
  {N{\"u}tzenadel}}, \bibinfo {author} {\bibfnamefont {S.}~\bibnamefont
  {Burgmann}}, \bibinfo {author} {\bibfnamefont {J.}~\bibnamefont {Czarske}},\
  and\ \bibinfo {author} {\bibfnamefont {L.}~\bibnamefont {B{\"u}ttner}},\
  }\bibfield  {title} {\bibinfo {title} {Adaptive-optical 3d microscopy for
  microfluidic multiphase flows},\ }\href@noop {} {\bibfield  {journal}
  {\bibinfo  {journal} {Light: Advanced Manufacturing}\ }\textbf {\bibinfo
  {volume} {5}},\ \bibinfo {pages} {385} (\bibinfo {year} {2024})}\BibitemShut
  {NoStop}%
\bibitem [{\citenamefont {Bertolotti}\ and\ \citenamefont
  {Katz}(2022)}]{bertolotti2022imaging}%
  \BibitemOpen
  \bibfield  {author} {\bibinfo {author} {\bibfnamefont {J.}~\bibnamefont
  {Bertolotti}}\ and\ \bibinfo {author} {\bibfnamefont {O.}~\bibnamefont
  {Katz}},\ }\bibfield  {title} {\bibinfo {title} {Imaging in complex media},\
  }\href@noop {} {\bibfield  {journal} {\bibinfo  {journal} {Nature Physics}\
  }\textbf {\bibinfo {volume} {18}},\ \bibinfo {pages} {1008} (\bibinfo {year}
  {2022})}\BibitemShut {NoStop}%
\bibitem [{\citenamefont {Wen}\ \emph {et~al.}(2023)\citenamefont {Wen},
  \citenamefont {Dong}, \citenamefont {Deng}, \citenamefont {Pang},
  \citenamefont {Kaminski}, \citenamefont {Xu}, \citenamefont {Yan},
  \citenamefont {Wang}, \citenamefont {Liu}, \citenamefont {Tang} \emph
  {et~al.}}]{wen2023single}%
  \BibitemOpen
  \bibfield  {author} {\bibinfo {author} {\bibfnamefont {Z.}~\bibnamefont
  {Wen}}, \bibinfo {author} {\bibfnamefont {Z.}~\bibnamefont {Dong}}, \bibinfo
  {author} {\bibfnamefont {Q.}~\bibnamefont {Deng}}, \bibinfo {author}
  {\bibfnamefont {C.}~\bibnamefont {Pang}}, \bibinfo {author} {\bibfnamefont
  {C.~F.}\ \bibnamefont {Kaminski}}, \bibinfo {author} {\bibfnamefont
  {X.}~\bibnamefont {Xu}}, \bibinfo {author} {\bibfnamefont {H.}~\bibnamefont
  {Yan}}, \bibinfo {author} {\bibfnamefont {L.}~\bibnamefont {Wang}}, \bibinfo
  {author} {\bibfnamefont {S.}~\bibnamefont {Liu}}, \bibinfo {author}
  {\bibfnamefont {J.}~\bibnamefont {Tang}}, \emph {et~al.},\ }\bibfield
  {title} {\bibinfo {title} {Single multimode fibre for in vivo
  light-field-encoded endoscopic imaging},\ }\href@noop {} {\bibfield
  {journal} {\bibinfo  {journal} {Nature Photonics}\ }\textbf {\bibinfo
  {volume} {17}},\ \bibinfo {pages} {679} (\bibinfo {year} {2023})}\BibitemShut
  {NoStop}%
\bibitem [{\citenamefont {Popoff}\ \emph
  {et~al.}(2010{\natexlab{a}})\citenamefont {Popoff}, \citenamefont {Lerosey},
  \citenamefont {Carminati}, \citenamefont {Fink}, \citenamefont {Boccara},\
  and\ \citenamefont {Gigan}}]{PhysRevLett2010measure}%
  \BibitemOpen
  \bibfield  {author} {\bibinfo {author} {\bibfnamefont {S.~M.}\ \bibnamefont
  {Popoff}}, \bibinfo {author} {\bibfnamefont {G.}~\bibnamefont {Lerosey}},
  \bibinfo {author} {\bibfnamefont {R.}~\bibnamefont {Carminati}}, \bibinfo
  {author} {\bibfnamefont {M.}~\bibnamefont {Fink}}, \bibinfo {author}
  {\bibfnamefont {A.~C.}\ \bibnamefont {Boccara}},\ and\ \bibinfo {author}
  {\bibfnamefont {S.}~\bibnamefont {Gigan}},\ }\bibfield  {title} {\bibinfo
  {title} {Measuring the transmission matrix in optics: An approach to the
  study and control of light propagation in disordered media},\ }\href@noop {}
  {\bibfield  {journal} {\bibinfo  {journal} {Physical Review Letters}\
  }\textbf {\bibinfo {volume} {104}},\ \bibinfo {pages} {100601} (\bibinfo
  {year} {2010}{\natexlab{a}})}\BibitemShut {NoStop}%
\bibitem [{\citenamefont {Popoff}\ \emph
  {et~al.}(2010{\natexlab{b}})\citenamefont {Popoff}, \citenamefont {Lerosey},
  \citenamefont {Fink}, \citenamefont {Boccara},\ and\ \citenamefont
  {Gigan}}]{popoff2010image}%
  \BibitemOpen
  \bibfield  {author} {\bibinfo {author} {\bibfnamefont {S.}~\bibnamefont
  {Popoff}}, \bibinfo {author} {\bibfnamefont {G.}~\bibnamefont {Lerosey}},
  \bibinfo {author} {\bibfnamefont {M.}~\bibnamefont {Fink}}, \bibinfo {author}
  {\bibfnamefont {A.~C.}\ \bibnamefont {Boccara}},\ and\ \bibinfo {author}
  {\bibfnamefont {S.}~\bibnamefont {Gigan}},\ }\bibfield  {title} {\bibinfo
  {title} {Image transmission through an opaque material},\ }\href@noop {}
  {\bibfield  {journal} {\bibinfo  {journal} {Nature communications}\ }\textbf
  {\bibinfo {volume} {1}},\ \bibinfo {pages} {81} (\bibinfo {year}
  {2010}{\natexlab{b}})}\BibitemShut {NoStop}%
\bibitem [{\citenamefont {Boniface}\ \emph {et~al.}(2020)\citenamefont
  {Boniface}, \citenamefont {Dong},\ and\ \citenamefont
  {Gigan}}]{boniface2020non}%
  \BibitemOpen
  \bibfield  {author} {\bibinfo {author} {\bibfnamefont {A.}~\bibnamefont
  {Boniface}}, \bibinfo {author} {\bibfnamefont {J.}~\bibnamefont {Dong}},\
  and\ \bibinfo {author} {\bibfnamefont {S.}~\bibnamefont {Gigan}},\ }\bibfield
   {title} {\bibinfo {title} {Non-invasive focusing and imaging in scattering
  media with a fluorescence-based transmission matrix},\ }\href@noop {}
  {\bibfield  {journal} {\bibinfo  {journal} {Nature communications}\ }\textbf
  {\bibinfo {volume} {11}},\ \bibinfo {pages} {6154} (\bibinfo {year}
  {2020})}\BibitemShut {NoStop}%
\bibitem [{\citenamefont {Yaqoob}\ \emph {et~al.}(2008)\citenamefont {Yaqoob},
  \citenamefont {Psaltis}, \citenamefont {Feld},\ and\ \citenamefont
  {Yang}}]{yaqoob2008optical}%
  \BibitemOpen
  \bibfield  {author} {\bibinfo {author} {\bibfnamefont {Z.}~\bibnamefont
  {Yaqoob}}, \bibinfo {author} {\bibfnamefont {D.}~\bibnamefont {Psaltis}},
  \bibinfo {author} {\bibfnamefont {M.~S.}\ \bibnamefont {Feld}},\ and\
  \bibinfo {author} {\bibfnamefont {C.}~\bibnamefont {Yang}},\ }\bibfield
  {title} {\bibinfo {title} {Optical phase conjugation for turbidity
  suppression in biological samples},\ }\href@noop {} {\bibfield  {journal}
  {\bibinfo  {journal} {Nature photonics}\ }\textbf {\bibinfo {volume} {2}},\
  \bibinfo {pages} {110} (\bibinfo {year} {2008})}\BibitemShut {NoStop}%
\bibitem [{\citenamefont {Xu}\ \emph {et~al.}(2011)\citenamefont {Xu},
  \citenamefont {Liu},\ and\ \citenamefont {Wang}}]{xu2011time}%
  \BibitemOpen
  \bibfield  {author} {\bibinfo {author} {\bibfnamefont {X.}~\bibnamefont
  {Xu}}, \bibinfo {author} {\bibfnamefont {H.}~\bibnamefont {Liu}},\ and\
  \bibinfo {author} {\bibfnamefont {L.~V.}\ \bibnamefont {Wang}},\ }\bibfield
  {title} {\bibinfo {title} {Time-reversed ultrasonically encoded optical
  focusing into scattering media},\ }\href@noop {} {\bibfield  {journal}
  {\bibinfo  {journal} {Nature photonics}\ }\textbf {\bibinfo {volume} {5}},\
  \bibinfo {pages} {154} (\bibinfo {year} {2011})}\BibitemShut {NoStop}%
\bibitem [{\citenamefont {Mosk}\ \emph {et~al.}(2012)\citenamefont {Mosk},
  \citenamefont {Lagendijk}, \citenamefont {Lerosey},\ and\ \citenamefont
  {Fink}}]{mosk2012controlling}%
  \BibitemOpen
  \bibfield  {author} {\bibinfo {author} {\bibfnamefont {A.~P.}\ \bibnamefont
  {Mosk}}, \bibinfo {author} {\bibfnamefont {A.}~\bibnamefont {Lagendijk}},
  \bibinfo {author} {\bibfnamefont {G.}~\bibnamefont {Lerosey}},\ and\ \bibinfo
  {author} {\bibfnamefont {M.}~\bibnamefont {Fink}},\ }\bibfield  {title}
  {\bibinfo {title} {Controlling waves in space and time for imaging and
  focusing in complex media},\ }\href@noop {} {\bibfield  {journal} {\bibinfo
  {journal} {Nature photonics}\ }\textbf {\bibinfo {volume} {6}},\ \bibinfo
  {pages} {283} (\bibinfo {year} {2012})}\BibitemShut {NoStop}%
\bibitem [{\citenamefont {Katz}\ \emph {et~al.}(2012)\citenamefont {Katz},
  \citenamefont {Small},\ and\ \citenamefont {Silberberg}}]{katz2012looking}%
  \BibitemOpen
  \bibfield  {author} {\bibinfo {author} {\bibfnamefont {O.}~\bibnamefont
  {Katz}}, \bibinfo {author} {\bibfnamefont {E.}~\bibnamefont {Small}},\ and\
  \bibinfo {author} {\bibfnamefont {Y.}~\bibnamefont {Silberberg}},\ }\bibfield
   {title} {\bibinfo {title} {Looking around corners and through thin turbid
  layers in real time with scattered incoherent light},\ }\href@noop {}
  {\bibfield  {journal} {\bibinfo  {journal} {Nature photonics}\ }\textbf
  {\bibinfo {volume} {6}},\ \bibinfo {pages} {549} (\bibinfo {year}
  {2012})}\BibitemShut {NoStop}%
\bibitem [{\citenamefont {Yeminy}\ and\ \citenamefont
  {Katz}(2021)}]{sciadv2021Tomer}%
  \BibitemOpen
  \bibfield  {author} {\bibinfo {author} {\bibfnamefont {T.}~\bibnamefont
  {Yeminy}}\ and\ \bibinfo {author} {\bibfnamefont {O.}~\bibnamefont {Katz}},\
  }\bibfield  {title} {\bibinfo {title} {Guidestar-free image-guided wavefront
  shaping},\ }\href@noop {} {\bibfield  {journal} {\bibinfo  {journal} {Science
  Advances}\ }\textbf {\bibinfo {volume} {7}},\ \bibinfo {pages} {eabf5364}
  (\bibinfo {year} {2021})}\BibitemShut {NoStop}%
\bibitem [{\citenamefont {Sun}\ \emph {et~al.}(2024)\citenamefont {Sun},
  \citenamefont {Nie}, \citenamefont {Du}, \citenamefont {Chang},\ and\
  \citenamefont {Liu}}]{Sun2024optica}%
  \BibitemOpen
  \bibfield  {author} {\bibinfo {author} {\bibfnamefont {S.}~\bibnamefont
  {Sun}}, \bibinfo {author} {\bibfnamefont {Z.-W.}\ \bibnamefont {Nie}},
  \bibinfo {author} {\bibfnamefont {L.-K.}\ \bibnamefont {Du}}, \bibinfo
  {author} {\bibfnamefont {C.}~\bibnamefont {Chang}},\ and\ \bibinfo {author}
  {\bibfnamefont {W.-T.}\ \bibnamefont {Liu}},\ }\bibfield  {title} {\bibinfo
  {title} {Overcoming the diffraction limit by exploiting unmeasured scattering
  media},\ }\href@noop {} {\bibfield  {journal} {\bibinfo  {journal} {Optica}\
  }\textbf {\bibinfo {volume} {11}},\ \bibinfo {pages} {385} (\bibinfo {year}
  {2024})}\BibitemShut {NoStop}%
\bibitem [{\citenamefont {Bertolotti}\ \emph {et~al.}(2012)\citenamefont
  {Bertolotti}, \citenamefont {Van~Putten}, \citenamefont {Blum}, \citenamefont
  {Lagendijk}, \citenamefont {Vos},\ and\ \citenamefont
  {Mosk}}]{bertolotti2012non}%
  \BibitemOpen
  \bibfield  {author} {\bibinfo {author} {\bibfnamefont {J.}~\bibnamefont
  {Bertolotti}}, \bibinfo {author} {\bibfnamefont {E.~G.}\ \bibnamefont
  {Van~Putten}}, \bibinfo {author} {\bibfnamefont {C.}~\bibnamefont {Blum}},
  \bibinfo {author} {\bibfnamefont {A.}~\bibnamefont {Lagendijk}}, \bibinfo
  {author} {\bibfnamefont {W.~L.}\ \bibnamefont {Vos}},\ and\ \bibinfo {author}
  {\bibfnamefont {A.~P.}\ \bibnamefont {Mosk}},\ }\bibfield  {title} {\bibinfo
  {title} {Non-invasive imaging through opaque scattering layers},\ }\href@noop
  {} {\bibfield  {journal} {\bibinfo  {journal} {Nature}\ }\textbf {\bibinfo
  {volume} {491}},\ \bibinfo {pages} {232} (\bibinfo {year}
  {2012})}\BibitemShut {NoStop}%
\bibitem [{\citenamefont {Katz}\ \emph {et~al.}(2014)\citenamefont {Katz},
  \citenamefont {Heidmann}, \citenamefont {Fink},\ and\ \citenamefont
  {Gigan}}]{katz2014non}%
  \BibitemOpen
  \bibfield  {author} {\bibinfo {author} {\bibfnamefont {O.}~\bibnamefont
  {Katz}}, \bibinfo {author} {\bibfnamefont {P.}~\bibnamefont {Heidmann}},
  \bibinfo {author} {\bibfnamefont {M.}~\bibnamefont {Fink}},\ and\ \bibinfo
  {author} {\bibfnamefont {S.}~\bibnamefont {Gigan}},\ }\bibfield  {title}
  {\bibinfo {title} {Non-invasive single-shot imaging through scattering layers
  and around corners via speckle correlations},\ }\href@noop {} {\bibfield
  {journal} {\bibinfo  {journal} {Nature photonics}\ }\textbf {\bibinfo
  {volume} {8}},\ \bibinfo {pages} {784} (\bibinfo {year} {2014})}\BibitemShut
  {NoStop}%
\bibitem [{\citenamefont {Zhu}\ \emph {et~al.}(2022)\citenamefont {Zhu},
  \citenamefont {Soldevila}, \citenamefont {Moretti}, \citenamefont {d’Arco},
  \citenamefont {Boniface}, \citenamefont {Shao}, \citenamefont {de~Aguiar},\
  and\ \citenamefont {Gigan}}]{zhu2022large}%
  \BibitemOpen
  \bibfield  {author} {\bibinfo {author} {\bibfnamefont {L.}~\bibnamefont
  {Zhu}}, \bibinfo {author} {\bibfnamefont {F.}~\bibnamefont {Soldevila}},
  \bibinfo {author} {\bibfnamefont {C.}~\bibnamefont {Moretti}}, \bibinfo
  {author} {\bibfnamefont {A.}~\bibnamefont {d’Arco}}, \bibinfo {author}
  {\bibfnamefont {A.}~\bibnamefont {Boniface}}, \bibinfo {author}
  {\bibfnamefont {X.}~\bibnamefont {Shao}}, \bibinfo {author} {\bibfnamefont
  {H.~B.}\ \bibnamefont {de~Aguiar}},\ and\ \bibinfo {author} {\bibfnamefont
  {S.}~\bibnamefont {Gigan}},\ }\bibfield  {title} {\bibinfo {title} {Large
  field-of-view non-invasive imaging through scattering layers using
  fluctuating random illumination},\ }\href@noop {} {\bibfield  {journal}
  {\bibinfo  {journal} {Nature communications}\ }\textbf {\bibinfo {volume}
  {13}},\ \bibinfo {pages} {1447} (\bibinfo {year} {2022})}\BibitemShut
  {NoStop}%
\bibitem [{\citenamefont {Li}\ \emph {et~al.}(2018{\natexlab{a}})\citenamefont
  {Li}, \citenamefont {Deng}, \citenamefont {Lee}, \citenamefont {Sinha},\ and\
  \citenamefont {Barbastathis}}]{Li2018imaging}%
  \BibitemOpen
  \bibfield  {author} {\bibinfo {author} {\bibfnamefont {S.}~\bibnamefont
  {Li}}, \bibinfo {author} {\bibfnamefont {M.}~\bibnamefont {Deng}}, \bibinfo
  {author} {\bibfnamefont {J.}~\bibnamefont {Lee}}, \bibinfo {author}
  {\bibfnamefont {A.}~\bibnamefont {Sinha}},\ and\ \bibinfo {author}
  {\bibfnamefont {G.}~\bibnamefont {Barbastathis}},\ }\bibfield  {title}
  {\bibinfo {title} {Imaging through glass diffusers using densely connected
  convolutional networks},\ }\href@noop {} {\bibfield  {journal} {\bibinfo
  {journal} {Optica}\ }\textbf {\bibinfo {volume} {5}},\ \bibinfo {pages} {803}
  (\bibinfo {year} {2018}{\natexlab{a}})}\BibitemShut {NoStop}%
\bibitem [{\citenamefont {Li}\ \emph {et~al.}(2018{\natexlab{b}})\citenamefont
  {Li}, \citenamefont {Xue},\ and\ \citenamefont {Tian}}]{Li2018deep}%
  \BibitemOpen
  \bibfield  {author} {\bibinfo {author} {\bibfnamefont {Y.}~\bibnamefont
  {Li}}, \bibinfo {author} {\bibfnamefont {Y.}~\bibnamefont {Xue}},\ and\
  \bibinfo {author} {\bibfnamefont {L.}~\bibnamefont {Tian}},\ }\bibfield
  {title} {\bibinfo {title} {Deep speckle correlation: a deep learning approach
  toward scalable imaging through scattering media},\ }\href@noop {} {\bibfield
   {journal} {\bibinfo  {journal} {Optica}\ }\textbf {\bibinfo {volume} {5}},\
  \bibinfo {pages} {1181} (\bibinfo {year} {2018}{\natexlab{b}})}\BibitemShut
  {NoStop}%
\bibitem [{\citenamefont {Zhu}\ \emph {et~al.}(2021)\citenamefont {Zhu},
  \citenamefont {Guo}, \citenamefont {Gu}, \citenamefont {Bai},\ and\
  \citenamefont {Han}}]{Zhu2021}%
  \BibitemOpen
  \bibfield  {author} {\bibinfo {author} {\bibfnamefont {S.}~\bibnamefont
  {Zhu}}, \bibinfo {author} {\bibfnamefont {E.}~\bibnamefont {Guo}}, \bibinfo
  {author} {\bibfnamefont {J.}~\bibnamefont {Gu}}, \bibinfo {author}
  {\bibfnamefont {L.}~\bibnamefont {Bai}},\ and\ \bibinfo {author}
  {\bibfnamefont {J.}~\bibnamefont {Han}},\ }\bibfield  {title} {\bibinfo
  {title} {Imaging through unknown scattering media based on physics-informed
  learning},\ }\href@noop {} {\bibfield  {journal} {\bibinfo  {journal}
  {Photonics Research.}\ }\textbf {\bibinfo {volume} {9}},\ \bibinfo {pages}
  {B210} (\bibinfo {year} {2021})}\BibitemShut {NoStop}%
\bibitem [{\citenamefont {Lyu}\ \emph {et~al.}(2019)\citenamefont {Lyu},
  \citenamefont {Wang}, \citenamefont {Li}, \citenamefont {Zheng},\ and\
  \citenamefont {Situ}}]{Mengap2019}%
  \BibitemOpen
  \bibfield  {author} {\bibinfo {author} {\bibfnamefont {M.}~\bibnamefont
  {Lyu}}, \bibinfo {author} {\bibfnamefont {H.}~\bibnamefont {Wang}}, \bibinfo
  {author} {\bibfnamefont {G.}~\bibnamefont {Li}}, \bibinfo {author}
  {\bibfnamefont {S.}~\bibnamefont {Zheng}},\ and\ \bibinfo {author}
  {\bibfnamefont {G.}~\bibnamefont {Situ}},\ }\bibfield  {title} {\bibinfo
  {title} {{Learning-based lensless imaging through optically thick scattering
  media}},\ }\href@noop {} {\bibfield  {journal} {\bibinfo  {journal} {Advanced
  Photonics}\ }\textbf {\bibinfo {volume} {1}},\ \bibinfo {pages} {036002}
  (\bibinfo {year} {2019})}\BibitemShut {NoStop}%
\bibitem [{\citenamefont {Zhang}\ \emph {et~al.}(2023)\citenamefont {Zhang},
  \citenamefont {Cheng}, \citenamefont {Gao}, \citenamefont {Gan},
  \citenamefont {Song}, \citenamefont {Zhang}, \citenamefont {Zhuang},
  \citenamefont {Han}, \citenamefont {Lai},\ and\ \citenamefont
  {Liu}}]{Zhang2023}%
  \BibitemOpen
  \bibfield  {author} {\bibinfo {author} {\bibfnamefont {X.}~\bibnamefont
  {Zhang}}, \bibinfo {author} {\bibfnamefont {S.}~\bibnamefont {Cheng}},
  \bibinfo {author} {\bibfnamefont {J.}~\bibnamefont {Gao}}, \bibinfo {author}
  {\bibfnamefont {Y.}~\bibnamefont {Gan}}, \bibinfo {author} {\bibfnamefont
  {C.}~\bibnamefont {Song}}, \bibinfo {author} {\bibfnamefont {D.}~\bibnamefont
  {Zhang}}, \bibinfo {author} {\bibfnamefont {S.}~\bibnamefont {Zhuang}},
  \bibinfo {author} {\bibfnamefont {S.}~\bibnamefont {Han}}, \bibinfo {author}
  {\bibfnamefont {P.}~\bibnamefont {Lai}},\ and\ \bibinfo {author}
  {\bibfnamefont {H.}~\bibnamefont {Liu}},\ }\bibfield  {title} {\bibinfo
  {title} {Physical origin and boundary of scalable imaging through scattering
  media: a deep learning-based exploration},\ }\href@noop {} {\bibfield
  {journal} {\bibinfo  {journal} {Photonics Research}\ }\textbf {\bibinfo
  {volume} {11}},\ \bibinfo {pages} {1038} (\bibinfo {year}
  {2023})}\BibitemShut {NoStop}%
\bibitem [{\citenamefont {Gao}\ \emph {et~al.}(2021)\citenamefont {Gao},
  \citenamefont {Radner}, \citenamefont {B{\"u}ttner}, \citenamefont {Ye},
  \citenamefont {Li},\ and\ \citenamefont {Czarske}}]{gao2021distortion}%
  \BibitemOpen
  \bibfield  {author} {\bibinfo {author} {\bibfnamefont {Z.}~\bibnamefont
  {Gao}}, \bibinfo {author} {\bibfnamefont {H.}~\bibnamefont {Radner}},
  \bibinfo {author} {\bibfnamefont {L.}~\bibnamefont {B{\"u}ttner}}, \bibinfo
  {author} {\bibfnamefont {H.}~\bibnamefont {Ye}}, \bibinfo {author}
  {\bibfnamefont {X.}~\bibnamefont {Li}},\ and\ \bibinfo {author}
  {\bibfnamefont {J.}~\bibnamefont {Czarske}},\ }\bibfield  {title} {\bibinfo
  {title} {Distortion correction for particle image velocimetry using
  multiple-input deep convolutional neural network and hartmann-shack
  sensing},\ }\href@noop {} {\bibfield  {journal} {\bibinfo  {journal} {Optics
  Express}\ }\textbf {\bibinfo {volume} {29}},\ \bibinfo {pages} {18669}
  (\bibinfo {year} {2021})}\BibitemShut {NoStop}%
\bibitem [{\citenamefont {Yang}\ \emph {et~al.}(2021)\citenamefont {Yang},
  \citenamefont {He}, \citenamefont {Liu}, \citenamefont {Qu}, \citenamefont
  {Shao}, \citenamefont {Song},\ and\ \citenamefont {Zhao}}]{yang2021anti}%
  \BibitemOpen
  \bibfield  {author} {\bibinfo {author} {\bibfnamefont {J.}~\bibnamefont
  {Yang}}, \bibinfo {author} {\bibfnamefont {Q.}~\bibnamefont {He}}, \bibinfo
  {author} {\bibfnamefont {L.}~\bibnamefont {Liu}}, \bibinfo {author}
  {\bibfnamefont {Y.}~\bibnamefont {Qu}}, \bibinfo {author} {\bibfnamefont
  {R.}~\bibnamefont {Shao}}, \bibinfo {author} {\bibfnamefont {B.}~\bibnamefont
  {Song}},\ and\ \bibinfo {author} {\bibfnamefont {Y.}~\bibnamefont {Zhao}},\
  }\bibfield  {title} {\bibinfo {title} {Anti-scattering light focusing by fast
  wavefront shaping based on multi-pixel encoded digital-micromirror device},\
  }\href@noop {} {\bibfield  {journal} {\bibinfo  {journal} {Light: Science \&
  Applications}\ }\textbf {\bibinfo {volume} {10}},\ \bibinfo {pages} {149}
  (\bibinfo {year} {2021})}\BibitemShut {NoStop}%
\bibitem [{\citenamefont {Tzang}\ \emph {et~al.}(2019)\citenamefont {Tzang},
  \citenamefont {Niv}, \citenamefont {Singh}, \citenamefont {Labouesse},
  \citenamefont {Myatt},\ and\ \citenamefont {Piestun}}]{tzang2019wavefront}%
  \BibitemOpen
  \bibfield  {author} {\bibinfo {author} {\bibfnamefont {O.}~\bibnamefont
  {Tzang}}, \bibinfo {author} {\bibfnamefont {E.}~\bibnamefont {Niv}}, \bibinfo
  {author} {\bibfnamefont {S.}~\bibnamefont {Singh}}, \bibinfo {author}
  {\bibfnamefont {S.}~\bibnamefont {Labouesse}}, \bibinfo {author}
  {\bibfnamefont {G.}~\bibnamefont {Myatt}},\ and\ \bibinfo {author}
  {\bibfnamefont {R.}~\bibnamefont {Piestun}},\ }\bibfield  {title} {\bibinfo
  {title} {Wavefront shaping in complex media with a 350 khz modulator via a
  1d-to-2d transform},\ }\href@noop {} {\bibfield  {journal} {\bibinfo
  {journal} {Nature Photonics}\ }\textbf {\bibinfo {volume} {13}},\ \bibinfo
  {pages} {788} (\bibinfo {year} {2019})}\BibitemShut {NoStop}%
\bibitem [{\citenamefont {Valzania}\ and\ \citenamefont
  {Gigan}(2023)}]{valzania2023online}%
  \BibitemOpen
  \bibfield  {author} {\bibinfo {author} {\bibfnamefont {L.}~\bibnamefont
  {Valzania}}\ and\ \bibinfo {author} {\bibfnamefont {S.}~\bibnamefont
  {Gigan}},\ }\bibfield  {title} {\bibinfo {title} {Online learning of the
  transmission matrix of dynamic scattering media},\ }\href@noop {} {\bibfield
  {journal} {\bibinfo  {journal} {Optica}\ }\textbf {\bibinfo {volume} {10}},\
  \bibinfo {pages} {708} (\bibinfo {year} {2023})}\BibitemShut {NoStop}%
\bibitem [{\citenamefont {Feng}\ \emph {et~al.}(2023)\citenamefont {Feng},
  \citenamefont {Guo}, \citenamefont {Xie}, \citenamefont {Boominathan},
  \citenamefont {Sharma}, \citenamefont {Veeraraghavan},\ and\ \citenamefont
  {Metzler}}]{sciadvBrandon2023}%
  \BibitemOpen
  \bibfield  {author} {\bibinfo {author} {\bibfnamefont {B.~Y.}\ \bibnamefont
  {Feng}}, \bibinfo {author} {\bibfnamefont {H.}~\bibnamefont {Guo}}, \bibinfo
  {author} {\bibfnamefont {M.}~\bibnamefont {Xie}}, \bibinfo {author}
  {\bibfnamefont {V.}~\bibnamefont {Boominathan}}, \bibinfo {author}
  {\bibfnamefont {M.~K.}\ \bibnamefont {Sharma}}, \bibinfo {author}
  {\bibfnamefont {A.}~\bibnamefont {Veeraraghavan}},\ and\ \bibinfo {author}
  {\bibfnamefont {C.~A.}\ \bibnamefont {Metzler}},\ }\bibfield  {title}
  {\bibinfo {title} {Neuws: Neural wavefront shaping for guidestar-free imaging
  through static and dynamic scattering media},\ }\href@noop {} {\bibfield
  {journal} {\bibinfo  {journal} {Science Advances}\ }\textbf {\bibinfo
  {volume} {9}},\ \bibinfo {pages} {eadg4671} (\bibinfo {year}
  {2023})}\BibitemShut {NoStop}%
\bibitem [{\citenamefont {He}\ \emph {et~al.}(2024)\citenamefont {He},
  \citenamefont {Shao}, \citenamefont {Qu}, \citenamefont {Liu}, \citenamefont
  {Ding},\ and\ \citenamefont {Yang}}]{He2024pr}%
  \BibitemOpen
  \bibfield  {author} {\bibinfo {author} {\bibfnamefont {Q.}~\bibnamefont
  {He}}, \bibinfo {author} {\bibfnamefont {R.}~\bibnamefont {Shao}}, \bibinfo
  {author} {\bibfnamefont {Y.}~\bibnamefont {Qu}}, \bibinfo {author}
  {\bibfnamefont {L.}~\bibnamefont {Liu}}, \bibinfo {author} {\bibfnamefont
  {C.}~\bibnamefont {Ding}},\ and\ \bibinfo {author} {\bibfnamefont
  {J.}~\bibnamefont {Yang}},\ }\bibfield  {title} {\bibinfo {title} {Complex
  transmission matrix retrieval for a highly scattering medium via regional
  phase differentiation},\ }\href@noop {} {\bibfield  {journal} {\bibinfo
  {journal} {Photonics Research}\ }\textbf {\bibinfo {volume} {12}},\ \bibinfo
  {pages} {876} (\bibinfo {year} {2024})}\BibitemShut {NoStop}%
\bibitem [{\citenamefont {Luo}\ \emph {et~al.}(2022{\natexlab{a}})\citenamefont
  {Luo}, \citenamefont {Liu}, \citenamefont {Wu}, \citenamefont {Xu},
  \citenamefont {Shao}, \citenamefont {Feng}, \citenamefont {Pan},
  \citenamefont {Zhao}, \citenamefont {Shen},\ and\ \citenamefont
  {Li}}]{sciadvadd9158}%
  \BibitemOpen
  \bibfield  {author} {\bibinfo {author} {\bibfnamefont {J.}~\bibnamefont
  {Luo}}, \bibinfo {author} {\bibfnamefont {Y.}~\bibnamefont {Liu}}, \bibinfo
  {author} {\bibfnamefont {D.}~\bibnamefont {Wu}}, \bibinfo {author}
  {\bibfnamefont {X.}~\bibnamefont {Xu}}, \bibinfo {author} {\bibfnamefont
  {L.}~\bibnamefont {Shao}}, \bibinfo {author} {\bibfnamefont {Y.}~\bibnamefont
  {Feng}}, \bibinfo {author} {\bibfnamefont {J.}~\bibnamefont {Pan}}, \bibinfo
  {author} {\bibfnamefont {J.}~\bibnamefont {Zhao}}, \bibinfo {author}
  {\bibfnamefont {Y.}~\bibnamefont {Shen}},\ and\ \bibinfo {author}
  {\bibfnamefont {Z.}~\bibnamefont {Li}},\ }\bibfield  {title} {\bibinfo
  {title} {High-speed single-exposure time-reversed ultrasonically encoded
  optical focusing against dynamic scattering},\ }\href@noop {} {\bibfield
  {journal} {\bibinfo  {journal} {Science Advances}\ }\textbf {\bibinfo
  {volume} {8}},\ \bibinfo {pages} {eadd9158} (\bibinfo {year}
  {2022}{\natexlab{a}})}\BibitemShut {NoStop}%
\bibitem [{\citenamefont {Liu}\ \emph {et~al.}(2015)\citenamefont {Liu},
  \citenamefont {Lai}, \citenamefont {Ma}, \citenamefont {Xu}, \citenamefont
  {Grabar},\ and\ \citenamefont {Wang}}]{liu2015optical}%
  \BibitemOpen
  \bibfield  {author} {\bibinfo {author} {\bibfnamefont {Y.}~\bibnamefont
  {Liu}}, \bibinfo {author} {\bibfnamefont {P.}~\bibnamefont {Lai}}, \bibinfo
  {author} {\bibfnamefont {C.}~\bibnamefont {Ma}}, \bibinfo {author}
  {\bibfnamefont {X.}~\bibnamefont {Xu}}, \bibinfo {author} {\bibfnamefont
  {A.~A.}\ \bibnamefont {Grabar}},\ and\ \bibinfo {author} {\bibfnamefont
  {L.~V.}\ \bibnamefont {Wang}},\ }\bibfield  {title} {\bibinfo {title}
  {Optical focusing deep inside dynamic scattering media with near-infrared
  time-reversed ultrasonically encoded (true) light},\ }\href@noop {}
  {\bibfield  {journal} {\bibinfo  {journal} {Nature communications}\ }\textbf
  {\bibinfo {volume} {6}},\ \bibinfo {pages} {5904} (\bibinfo {year}
  {2015})}\BibitemShut {NoStop}%
\bibitem [{\citenamefont {Liu}\ \emph {et~al.}(2017)\citenamefont {Liu},
  \citenamefont {Ma}, \citenamefont {Shen}, \citenamefont {Shi},\ and\
  \citenamefont {Wang}}]{Liu17optica}%
  \BibitemOpen
  \bibfield  {author} {\bibinfo {author} {\bibfnamefont {Y.}~\bibnamefont
  {Liu}}, \bibinfo {author} {\bibfnamefont {C.}~\bibnamefont {Ma}}, \bibinfo
  {author} {\bibfnamefont {Y.}~\bibnamefont {Shen}}, \bibinfo {author}
  {\bibfnamefont {J.}~\bibnamefont {Shi}},\ and\ \bibinfo {author}
  {\bibfnamefont {L.~V.}\ \bibnamefont {Wang}},\ }\bibfield  {title} {\bibinfo
  {title} {Focusing light inside dynamic scattering media with millisecond
  digital optical phase conjugation},\ }\href@noop {} {\bibfield  {journal}
  {\bibinfo  {journal} {Optica}\ }\textbf {\bibinfo {volume} {4}},\ \bibinfo
  {pages} {280} (\bibinfo {year} {2017})}\BibitemShut {NoStop}%
\bibitem [{\citenamefont {Wang}\ \emph {et~al.}(2015)\citenamefont {Wang},
  \citenamefont {Zhou}, \citenamefont {Brake}, \citenamefont {Ruan},
  \citenamefont {Jang},\ and\ \citenamefont {Yang}}]{Wang15optica}%
  \BibitemOpen
  \bibfield  {author} {\bibinfo {author} {\bibfnamefont {D.}~\bibnamefont
  {Wang}}, \bibinfo {author} {\bibfnamefont {E.~H.}\ \bibnamefont {Zhou}},
  \bibinfo {author} {\bibfnamefont {J.}~\bibnamefont {Brake}}, \bibinfo
  {author} {\bibfnamefont {H.}~\bibnamefont {Ruan}}, \bibinfo {author}
  {\bibfnamefont {M.}~\bibnamefont {Jang}},\ and\ \bibinfo {author}
  {\bibfnamefont {C.}~\bibnamefont {Yang}},\ }\bibfield  {title} {\bibinfo
  {title} {Focusing through dynamic tissue with millisecond digital optical
  phase conjugation},\ }\href@noop {} {\bibfield  {journal} {\bibinfo
  {journal} {Optica}\ }\textbf {\bibinfo {volume} {2}},\ \bibinfo {pages} {728}
  (\bibinfo {year} {2015})}\BibitemShut {NoStop}%
\bibitem [{\citenamefont {Fu}\ \emph {et~al.}(2024)\citenamefont {Fu},
  \citenamefont {Wang}, \citenamefont {Tang}, \citenamefont {Bian},\ and\
  \citenamefont {Situ}}]{fu2024adaptive}%
  \BibitemOpen
  \bibfield  {author} {\bibinfo {author} {\bibfnamefont {Z.}~\bibnamefont
  {Fu}}, \bibinfo {author} {\bibfnamefont {F.}~\bibnamefont {Wang}}, \bibinfo
  {author} {\bibfnamefont {Z.}~\bibnamefont {Tang}}, \bibinfo {author}
  {\bibfnamefont {Y.}~\bibnamefont {Bian}},\ and\ \bibinfo {author}
  {\bibfnamefont {G.}~\bibnamefont {Situ}},\ }\bibfield  {title} {\bibinfo
  {title} {Adaptive imaging through dense dynamic scattering media using
  transfer learning},\ }\href@noop {} {\bibfield  {journal} {\bibinfo
  {journal} {Optics Express}\ }\textbf {\bibinfo {volume} {32}},\ \bibinfo
  {pages} {13688} (\bibinfo {year} {2024})}\BibitemShut {NoStop}%
\bibitem [{\citenamefont {Liu}\ \emph {et~al.}(2023)\citenamefont {Liu},
  \citenamefont {Hu}, \citenamefont {Chu}, \citenamefont {Liu},\ and\
  \citenamefont {Zhou}}]{liu2023deep}%
  \BibitemOpen
  \bibfield  {author} {\bibinfo {author} {\bibfnamefont {Y.}~\bibnamefont
  {Liu}}, \bibinfo {author} {\bibfnamefont {G.}~\bibnamefont {Hu}}, \bibinfo
  {author} {\bibfnamefont {X.}~\bibnamefont {Chu}}, \bibinfo {author}
  {\bibfnamefont {Z.}~\bibnamefont {Liu}},\ and\ \bibinfo {author}
  {\bibfnamefont {L.}~\bibnamefont {Zhou}},\ }\bibfield  {title} {\bibinfo
  {title} {Deep learning based coherent diffraction imaging of dynamic
  scattering media},\ }\href@noop {} {\bibfield  {journal} {\bibinfo  {journal}
  {Optics Express}\ }\textbf {\bibinfo {volume} {31}},\ \bibinfo {pages}
  {44410} (\bibinfo {year} {2023})}\BibitemShut {NoStop}%
\bibitem [{\citenamefont {Zhang}\ \emph
  {et~al.}(2024{\natexlab{a}})\citenamefont {Zhang}, \citenamefont {Huang},
  \citenamefont {Zhang}, \citenamefont {Zhuang}, \citenamefont {Han},
  \citenamefont {Lai},\ and\ \citenamefont {Liu}}]{zhang2024gen}%
  \BibitemOpen
  \bibfield  {author} {\bibinfo {author} {\bibfnamefont {X.}~\bibnamefont
  {Zhang}}, \bibinfo {author} {\bibfnamefont {H.}~\bibnamefont {Huang}},
  \bibinfo {author} {\bibfnamefont {D.}~\bibnamefont {Zhang}}, \bibinfo
  {author} {\bibfnamefont {S.}~\bibnamefont {Zhuang}}, \bibinfo {author}
  {\bibfnamefont {S.}~\bibnamefont {Han}}, \bibinfo {author} {\bibfnamefont
  {P.}~\bibnamefont {Lai}},\ and\ \bibinfo {author} {\bibfnamefont
  {H.}~\bibnamefont {Liu}},\ }\bibfield  {title} {\bibinfo {title}
  {Generalization vs. hallucination},\ }\href@noop {} {\bibfield  {journal}
  {\bibinfo  {journal} {arXiv preprint arXiv:2411.02893}\ } (\bibinfo {year}
  {2024}{\natexlab{a}})}\BibitemShut {NoStop}%
\bibitem [{\citenamefont {Liu}\ \emph {et~al.}(2024)\citenamefont {Liu},
  \citenamefont {Wang}, \citenamefont {Jin}, \citenamefont {Ma}, \citenamefont
  {Li}, \citenamefont {Bian},\ and\ \citenamefont {Situ}}]{liu2024learning}%
  \BibitemOpen
  \bibfield  {author} {\bibinfo {author} {\bibfnamefont {H.}~\bibnamefont
  {Liu}}, \bibinfo {author} {\bibfnamefont {F.}~\bibnamefont {Wang}}, \bibinfo
  {author} {\bibfnamefont {Y.}~\bibnamefont {Jin}}, \bibinfo {author}
  {\bibfnamefont {X.}~\bibnamefont {Ma}}, \bibinfo {author} {\bibfnamefont
  {S.}~\bibnamefont {Li}}, \bibinfo {author} {\bibfnamefont {Y.}~\bibnamefont
  {Bian}},\ and\ \bibinfo {author} {\bibfnamefont {G.}~\bibnamefont {Situ}},\
  }\bibfield  {title} {\bibinfo {title} {Learning-based real-time imaging
  through dynamic scattering media},\ }\href@noop {} {\bibfield  {journal}
  {\bibinfo  {journal} {Light: Science \& Applications}\ }\textbf {\bibinfo
  {volume} {13}},\ \bibinfo {pages} {194} (\bibinfo {year} {2024})}\BibitemShut
  {NoStop}%
\bibitem [{\citenamefont {Luo}\ \emph {et~al.}(2021)\citenamefont {Luo},
  \citenamefont {Yan}, \citenamefont {Li}, \citenamefont {Lai},\ and\
  \citenamefont {Zheng}}]{Luo2021}%
  \BibitemOpen
  \bibfield  {author} {\bibinfo {author} {\bibfnamefont {Y.}~\bibnamefont
  {Luo}}, \bibinfo {author} {\bibfnamefont {S.}~\bibnamefont {Yan}}, \bibinfo
  {author} {\bibfnamefont {H.}~\bibnamefont {Li}}, \bibinfo {author}
  {\bibfnamefont {P.}~\bibnamefont {Lai}},\ and\ \bibinfo {author}
  {\bibfnamefont {Y.}~\bibnamefont {Zheng}},\ }\bibfield  {title} {\bibinfo
  {title} {Towards smart optical focusing: deep learning-empowered dynamic
  wavefront shaping through nonstationary scattering media},\ }\href@noop {}
  {\bibfield  {journal} {\bibinfo  {journal} {Photonics Research}\ }\textbf
  {\bibinfo {volume} {9}},\ \bibinfo {pages} {B262} (\bibinfo {year}
  {2021})}\BibitemShut {NoStop}%
\bibitem [{\citenamefont {Li}\ \emph {et~al.}(2024)\citenamefont {Li},
  \citenamefont {Zhou}, \citenamefont {Zhou}, \citenamefont {Zhang},
  \citenamefont {Shi}, \citenamefont {Shen}, \citenamefont {Zhang},
  \citenamefont {Chi},\ and\ \citenamefont {Dai}}]{li2024self}%
  \BibitemOpen
  \bibfield  {author} {\bibinfo {author} {\bibfnamefont {Z.}~\bibnamefont
  {Li}}, \bibinfo {author} {\bibfnamefont {W.}~\bibnamefont {Zhou}}, \bibinfo
  {author} {\bibfnamefont {Z.}~\bibnamefont {Zhou}}, \bibinfo {author}
  {\bibfnamefont {S.}~\bibnamefont {Zhang}}, \bibinfo {author} {\bibfnamefont
  {J.}~\bibnamefont {Shi}}, \bibinfo {author} {\bibfnamefont {C.}~\bibnamefont
  {Shen}}, \bibinfo {author} {\bibfnamefont {J.}~\bibnamefont {Zhang}},
  \bibinfo {author} {\bibfnamefont {N.}~\bibnamefont {Chi}},\ and\ \bibinfo
  {author} {\bibfnamefont {Q.}~\bibnamefont {Dai}},\ }\bibfield  {title}
  {\bibinfo {title} {Self-supervised dynamic learning for long-term
  high-fidelity image transmission through unstabilized diffusive media},\
  }\href@noop {} {\bibfield  {journal} {\bibinfo  {journal} {Nature
  Communications}\ }\textbf {\bibinfo {volume} {15}},\ \bibinfo {pages} {1498}
  (\bibinfo {year} {2024})}\BibitemShut {NoStop}%
\bibitem [{\citenamefont {Edrei}\ and\ \citenamefont
  {Scarcelli}(2016)}]{Edrei2016}%
  \BibitemOpen
  \bibfield  {author} {\bibinfo {author} {\bibfnamefont {E.}~\bibnamefont
  {Edrei}}\ and\ \bibinfo {author} {\bibfnamefont {G.}~\bibnamefont
  {Scarcelli}},\ }\bibfield  {title} {\bibinfo {title} {Optical imaging through
  dynamic turbid media using the fourier-domain shower-curtain effect},\
  }\href@noop {} {\bibfield  {journal} {\bibinfo  {journal} {Optica}\ }\textbf
  {\bibinfo {volume} {3}},\ \bibinfo {pages} {71} (\bibinfo {year}
  {2016})}\BibitemShut {NoStop}%
\bibitem [{\citenamefont {Hwang}\ \emph {et~al.}(2019)\citenamefont {Hwang},
  \citenamefont {Woo},\ and\ \citenamefont {Park}}]{Hwang2019bispectrum}%
  \BibitemOpen
  \bibfield  {author} {\bibinfo {author} {\bibfnamefont {B.}~\bibnamefont
  {Hwang}}, \bibinfo {author} {\bibfnamefont {T.}~\bibnamefont {Woo}},\ and\
  \bibinfo {author} {\bibfnamefont {J.-H.}\ \bibnamefont {Park}},\ }\bibfield
  {title} {\bibinfo {title} {Fast diffraction-limited image recovery through
  turbulence via subsampled bispectrum analysis},\ }\href@noop {} {\bibfield
  {journal} {\bibinfo  {journal} {Optics Letter}\ }\textbf {\bibinfo {volume}
  {44}},\ \bibinfo {pages} {5985} (\bibinfo {year} {2019})}\BibitemShut
  {NoStop}%
\bibitem [{\citenamefont {Hwang}\ \emph {et~al.}(2023)\citenamefont {Hwang},
  \citenamefont {Woo}, \citenamefont {Ahn},\ and\ \citenamefont
  {Park}}]{Hwang2023coherent}%
  \BibitemOpen
  \bibfield  {author} {\bibinfo {author} {\bibfnamefont {B.}~\bibnamefont
  {Hwang}}, \bibinfo {author} {\bibfnamefont {T.}~\bibnamefont {Woo}}, \bibinfo
  {author} {\bibfnamefont {C.}~\bibnamefont {Ahn}},\ and\ \bibinfo {author}
  {\bibfnamefont {J.-H.}\ \bibnamefont {Park}},\ }\bibfield  {title} {\bibinfo
  {title} {Imaging through random media using coherent averaging},\ }\href@noop
  {} {\bibfield  {journal} {\bibinfo  {journal} {Laser \& Photonics Reviews}\
  }\textbf {\bibinfo {volume} {17}},\ \bibinfo {pages} {2200673} (\bibinfo
  {year} {2023})}\BibitemShut {NoStop}%
\bibitem [{\citenamefont {Luo}\ \emph {et~al.}(2025)\citenamefont {Luo},
  \citenamefont {Wang}, \citenamefont {He}, \citenamefont {Vinu}, \citenamefont
  {Luo}, \citenamefont {Pu},\ and\ \citenamefont {Chen}}]{luosingle}%
  \BibitemOpen
  \bibfield  {author} {\bibinfo {author} {\bibfnamefont {Y.}~\bibnamefont
  {Luo}}, \bibinfo {author} {\bibfnamefont {Z.}~\bibnamefont {Wang}}, \bibinfo
  {author} {\bibfnamefont {H.}~\bibnamefont {He}}, \bibinfo {author}
  {\bibfnamefont {R.~V.}\ \bibnamefont {Vinu}}, \bibinfo {author}
  {\bibfnamefont {S.}~\bibnamefont {Luo}}, \bibinfo {author} {\bibfnamefont
  {J.}~\bibnamefont {Pu}},\ and\ \bibinfo {author} {\bibfnamefont
  {Z.}~\bibnamefont {Chen}},\ }\bibfield  {title} {\bibinfo {title}
  {Single-shot non-invasive imaging through dynamic scattering media beyond the
  memory effect via virtual reference-based correlation holography},\
  }\href@noop {} {\bibfield  {journal} {\bibinfo  {journal} {Laser \& Photonics
  Reviews}\ }\textbf {\bibinfo {volume} {19}},\ \bibinfo {pages} {2400978}
  (\bibinfo {year} {2025})}\BibitemShut {NoStop}%
\bibitem [{\citenamefont {Ruan}\ \emph {et~al.}(2020)\citenamefont {Ruan},
  \citenamefont {Liu}, \citenamefont {Xu}, \citenamefont {Huang},\ and\
  \citenamefont {Yang}}]{ruan2020fluorescence}%
  \BibitemOpen
  \bibfield  {author} {\bibinfo {author} {\bibfnamefont {H.}~\bibnamefont
  {Ruan}}, \bibinfo {author} {\bibfnamefont {Y.}~\bibnamefont {Liu}}, \bibinfo
  {author} {\bibfnamefont {J.}~\bibnamefont {Xu}}, \bibinfo {author}
  {\bibfnamefont {Y.}~\bibnamefont {Huang}},\ and\ \bibinfo {author}
  {\bibfnamefont {C.}~\bibnamefont {Yang}},\ }\bibfield  {title} {\bibinfo
  {title} {Fluorescence imaging through dynamic scattering media with
  speckle-encoded ultrasound-modulated light correlation},\ }\href@noop {}
  {\bibfield  {journal} {\bibinfo  {journal} {Nature Photonics}\ }\textbf
  {\bibinfo {volume} {14}},\ \bibinfo {pages} {511} (\bibinfo {year}
  {2020})}\BibitemShut {NoStop}%
\bibitem [{\citenamefont {Wang}\ \emph {et~al.}(2021)\citenamefont {Wang},
  \citenamefont {Sahoo}, \citenamefont {Zhu}, \citenamefont {Adamo},\ and\
  \citenamefont {Dang}}]{wang2021nc}%
  \BibitemOpen
  \bibfield  {author} {\bibinfo {author} {\bibfnamefont {D.}~\bibnamefont
  {Wang}}, \bibinfo {author} {\bibfnamefont {S.~K.}\ \bibnamefont {Sahoo}},
  \bibinfo {author} {\bibfnamefont {X.}~\bibnamefont {Zhu}}, \bibinfo {author}
  {\bibfnamefont {G.}~\bibnamefont {Adamo}},\ and\ \bibinfo {author}
  {\bibfnamefont {C.}~\bibnamefont {Dang}},\ }\bibfield  {title} {\bibinfo
  {title} {Non-invasive super-resolution imaging through dynamic scattering
  media},\ }\href@noop {} {\bibfield  {journal} {\bibinfo  {journal} {Nature
  Communications}\ }\textbf {\bibinfo {volume} {12}},\ \bibinfo {pages} {3150}
  (\bibinfo {year} {2021})}\BibitemShut {NoStop}%
\bibitem [{\citenamefont {Jauregui-S{\'a}nchez}\ \emph
  {et~al.}(2022)\citenamefont {Jauregui-S{\'a}nchez}, \citenamefont {Penketh},\
  and\ \citenamefont {Bertolotti}}]{jauregui2022tracking}%
  \BibitemOpen
  \bibfield  {author} {\bibinfo {author} {\bibfnamefont {Y.}~\bibnamefont
  {Jauregui-S{\'a}nchez}}, \bibinfo {author} {\bibfnamefont {H.}~\bibnamefont
  {Penketh}},\ and\ \bibinfo {author} {\bibfnamefont {J.}~\bibnamefont
  {Bertolotti}},\ }\bibfield  {title} {\bibinfo {title} {Tracking moving
  objects through scattering media via speckle correlations},\ }\href@noop {}
  {\bibfield  {journal} {\bibinfo  {journal} {nature communications}\ }\textbf
  {\bibinfo {volume} {13}},\ \bibinfo {pages} {5779} (\bibinfo {year}
  {2022})}\BibitemShut {NoStop}%
\bibitem [{\citenamefont {Lin}\ \emph {et~al.}(2018)\citenamefont {Lin},
  \citenamefont {Rivenson}, \citenamefont {Yardimci}, \citenamefont {Veli},
  \citenamefont {Luo}, \citenamefont {Jarrahi},\ and\ \citenamefont
  {Ozcan}}]{sciencea2018}%
  \BibitemOpen
  \bibfield  {author} {\bibinfo {author} {\bibfnamefont {X.}~\bibnamefont
  {Lin}}, \bibinfo {author} {\bibfnamefont {Y.}~\bibnamefont {Rivenson}},
  \bibinfo {author} {\bibfnamefont {N.~T.}\ \bibnamefont {Yardimci}}, \bibinfo
  {author} {\bibfnamefont {M.}~\bibnamefont {Veli}}, \bibinfo {author}
  {\bibfnamefont {Y.}~\bibnamefont {Luo}}, \bibinfo {author} {\bibfnamefont
  {M.}~\bibnamefont {Jarrahi}},\ and\ \bibinfo {author} {\bibfnamefont
  {A.}~\bibnamefont {Ozcan}},\ }\bibfield  {title} {\bibinfo {title}
  {All-optical machine learning using diffractive deep neural networks},\
  }\href@noop {} {\bibfield  {journal} {\bibinfo  {journal} {Science}\ }\textbf
  {\bibinfo {volume} {361}},\ \bibinfo {pages} {1004} (\bibinfo {year}
  {2018})}\BibitemShut {NoStop}%
\bibitem [{\citenamefont {Veli}\ \emph {et~al.}(2021)\citenamefont {Veli},
  \citenamefont {Mengu}, \citenamefont {Yardimci}, \citenamefont {Luo},
  \citenamefont {Li}, \citenamefont {Rivenson}, \citenamefont {Jarrahi},\ and\
  \citenamefont {Ozcan}}]{veli2021terahertz}%
  \BibitemOpen
  \bibfield  {author} {\bibinfo {author} {\bibfnamefont {M.}~\bibnamefont
  {Veli}}, \bibinfo {author} {\bibfnamefont {D.}~\bibnamefont {Mengu}},
  \bibinfo {author} {\bibfnamefont {N.~T.}\ \bibnamefont {Yardimci}}, \bibinfo
  {author} {\bibfnamefont {Y.}~\bibnamefont {Luo}}, \bibinfo {author}
  {\bibfnamefont {J.}~\bibnamefont {Li}}, \bibinfo {author} {\bibfnamefont
  {Y.}~\bibnamefont {Rivenson}}, \bibinfo {author} {\bibfnamefont
  {M.}~\bibnamefont {Jarrahi}},\ and\ \bibinfo {author} {\bibfnamefont
  {A.}~\bibnamefont {Ozcan}},\ }\bibfield  {title} {\bibinfo {title} {Terahertz
  pulse shaping using diffractive surfaces},\ }\href@noop {} {\bibfield
  {journal} {\bibinfo  {journal} {Nature Communications}\ }\textbf {\bibinfo
  {volume} {12}},\ \bibinfo {pages} {37} (\bibinfo {year} {2021})}\BibitemShut
  {NoStop}%
\bibitem [{\citenamefont {Yang}\ \emph {et~al.}(2024)\citenamefont {Yang},
  \citenamefont {Rahman}, \citenamefont {Bai}, \citenamefont {Li},\ and\
  \citenamefont {Ozcan}}]{yang2024apn}%
  \BibitemOpen
  \bibfield  {author} {\bibinfo {author} {\bibfnamefont {X.}~\bibnamefont
  {Yang}}, \bibinfo {author} {\bibfnamefont {M.~S.~S.}\ \bibnamefont {Rahman}},
  \bibinfo {author} {\bibfnamefont {B.}~\bibnamefont {Bai}}, \bibinfo {author}
  {\bibfnamefont {J.}~\bibnamefont {Li}},\ and\ \bibinfo {author}
  {\bibfnamefont {A.}~\bibnamefont {Ozcan}},\ }\bibfield  {title} {\bibinfo
  {title} {{Complex-valued universal linear transformations and image
  encryption using spatially incoherent diffractive networks}},\ }\href@noop {}
  {\bibfield  {journal} {\bibinfo  {journal} {Advanced Photonics Nexus}\
  }\textbf {\bibinfo {volume} {3}},\ \bibinfo {pages} {016010} (\bibinfo {year}
  {2024})}\BibitemShut {NoStop}%
\bibitem [{\citenamefont {Luo}\ \emph {et~al.}(2019)\citenamefont {Luo},
  \citenamefont {Mengu}, \citenamefont {Yardimci}, \citenamefont {Rivenson},
  \citenamefont {Veli}, \citenamefont {Jarrahi},\ and\ \citenamefont
  {Ozcan}}]{luo2019design}%
  \BibitemOpen
  \bibfield  {author} {\bibinfo {author} {\bibfnamefont {Y.}~\bibnamefont
  {Luo}}, \bibinfo {author} {\bibfnamefont {D.}~\bibnamefont {Mengu}}, \bibinfo
  {author} {\bibfnamefont {N.~T.}\ \bibnamefont {Yardimci}}, \bibinfo {author}
  {\bibfnamefont {Y.}~\bibnamefont {Rivenson}}, \bibinfo {author}
  {\bibfnamefont {M.}~\bibnamefont {Veli}}, \bibinfo {author} {\bibfnamefont
  {M.}~\bibnamefont {Jarrahi}},\ and\ \bibinfo {author} {\bibfnamefont
  {A.}~\bibnamefont {Ozcan}},\ }\bibfield  {title} {\bibinfo {title} {Design of
  task-specific optical systems using broadband diffractive neural networks},\
  }\href@noop {} {\bibfield  {journal} {\bibinfo  {journal} {Light: Science \&
  Applications}\ }\textbf {\bibinfo {volume} {8}},\ \bibinfo {pages} {112}
  (\bibinfo {year} {2019})}\BibitemShut {NoStop}%
\bibitem [{\citenamefont {Luo}\ \emph {et~al.}(2022{\natexlab{b}})\citenamefont
  {Luo}, \citenamefont {Zhao}, \citenamefont {Li}, \citenamefont
  {{\c{C}}etinta{\c{s}}}, \citenamefont {Rivenson}, \citenamefont {Jarrahi},\
  and\ \citenamefont {Ozcan}}]{luo2022computational}%
  \BibitemOpen
  \bibfield  {author} {\bibinfo {author} {\bibfnamefont {Y.}~\bibnamefont
  {Luo}}, \bibinfo {author} {\bibfnamefont {Y.}~\bibnamefont {Zhao}}, \bibinfo
  {author} {\bibfnamefont {J.}~\bibnamefont {Li}}, \bibinfo {author}
  {\bibfnamefont {E.}~\bibnamefont {{\c{C}}etinta{\c{s}}}}, \bibinfo {author}
  {\bibfnamefont {Y.}~\bibnamefont {Rivenson}}, \bibinfo {author}
  {\bibfnamefont {M.}~\bibnamefont {Jarrahi}},\ and\ \bibinfo {author}
  {\bibfnamefont {A.}~\bibnamefont {Ozcan}},\ }\bibfield  {title} {\bibinfo
  {title} {Computational imaging without a computer: seeing through random
  diffusers at the speed of light},\ }\href@noop {} {\bibfield  {journal}
  {\bibinfo  {journal} {ELight}\ }\textbf {\bibinfo {volume} {2}},\ \bibinfo
  {pages} {4} (\bibinfo {year} {2022}{\natexlab{b}})}\BibitemShut {NoStop}%
\bibitem [{\citenamefont {Li}\ \emph {et~al.}(2023{\natexlab{a}})\citenamefont
  {Li}, \citenamefont {Luo}, \citenamefont {Mengu}, \citenamefont {Bai},\ and\
  \citenamefont {Aydogan}}]{LAM2023010005}%
  \BibitemOpen
  \bibfield  {author} {\bibinfo {author} {\bibfnamefont {Y.}~\bibnamefont
  {Li}}, \bibinfo {author} {\bibfnamefont {Y.}~\bibnamefont {Luo}}, \bibinfo
  {author} {\bibfnamefont {D.}~\bibnamefont {Mengu}}, \bibinfo {author}
  {\bibfnamefont {B.}~\bibnamefont {Bai}},\ and\ \bibinfo {author}
  {\bibfnamefont {O.}~\bibnamefont {Aydogan}},\ }\bibfield  {title} {\bibinfo
  {title} {Quantitative phase imaging (qpi) through random diffusers using a
  diffractive optical network},\ }\href@noop {} {\bibfield  {journal} {\bibinfo
   {journal} {Light: Advanced Manufacturing}\ }\textbf {\bibinfo {volume}
  {4}},\ \bibinfo {pages} {206} (\bibinfo {year}
  {2023}{\natexlab{a}})}\BibitemShut {NoStop}%
\bibitem [{\citenamefont {Li}\ \emph {et~al.}(2023{\natexlab{b}})\citenamefont
  {Li}, \citenamefont {Gan}, \citenamefont {Bai}, \citenamefont {Işıl},
  \citenamefont {Jarrahi},\ and\ \citenamefont {Ozcan}}]{li2023opt}%
  \BibitemOpen
  \bibfield  {author} {\bibinfo {author} {\bibfnamefont {Y.}~\bibnamefont
  {Li}}, \bibinfo {author} {\bibfnamefont {T.}~\bibnamefont {Gan}}, \bibinfo
  {author} {\bibfnamefont {B.}~\bibnamefont {Bai}}, \bibinfo {author}
  {\bibfnamefont {{\c{C}}.}~\bibnamefont {Işıl}}, \bibinfo {author}
  {\bibfnamefont {M.}~\bibnamefont {Jarrahi}},\ and\ \bibinfo {author}
  {\bibfnamefont {A.}~\bibnamefont {Ozcan}},\ }\bibfield  {title} {\bibinfo
  {title} {{Optical information transfer through random unknown diffusers using
  electronic encoding and diffractive decoding}},\ }\href@noop {} {\bibfield
  {journal} {\bibinfo  {journal} {Advanced Photonics}\ }\textbf {\bibinfo
  {volume} {5}},\ \bibinfo {pages} {046009} (\bibinfo {year}
  {2023}{\natexlab{b}})}\BibitemShut {NoStop}%
\bibitem [{\citenamefont {Zhang}\ \emph
  {et~al.}(2024{\natexlab{b}})\citenamefont {Zhang}, \citenamefont {Zhang},
  \citenamefont {Yu}, \citenamefont {Zhang}, \citenamefont {Luan},\ and\
  \citenamefont {Gu}}]{sciadvadn2024zhang}%
  \BibitemOpen
  \bibfield  {author} {\bibinfo {author} {\bibfnamefont {Y.}~\bibnamefont
  {Zhang}}, \bibinfo {author} {\bibfnamefont {Q.}~\bibnamefont {Zhang}},
  \bibinfo {author} {\bibfnamefont {H.}~\bibnamefont {Yu}}, \bibinfo {author}
  {\bibfnamefont {Y.}~\bibnamefont {Zhang}}, \bibinfo {author} {\bibfnamefont
  {H.}~\bibnamefont {Luan}},\ and\ \bibinfo {author} {\bibfnamefont
  {M.}~\bibnamefont {Gu}},\ }\bibfield  {title} {\bibinfo {title} {Memory-less
  scattering imaging with ultrafast convolutional optical neural networks},\
  }\href@noop {} {\bibfield  {journal} {\bibinfo  {journal} {Science Advances}\
  }\textbf {\bibinfo {volume} {10}},\ \bibinfo {pages} {eadn2205} (\bibinfo
  {year} {2024}{\natexlab{b}})}\BibitemShut {NoStop}%
\bibitem [{\citenamefont {Yu}\ \emph {et~al.}(2025)\citenamefont {Yu},
  \citenamefont {Huang}, \citenamefont {Lamon}, \citenamefont {Wang},
  \citenamefont {Ding}, \citenamefont {Lin}, \citenamefont {Wang},
  \citenamefont {Luan}, \citenamefont {Gu},\ and\ \citenamefont
  {Zhang}}]{NP2025gumin}%
  \BibitemOpen
  \bibfield  {author} {\bibinfo {author} {\bibfnamefont {H.}~\bibnamefont
  {Yu}}, \bibinfo {author} {\bibfnamefont {Z.}~\bibnamefont {Huang}}, \bibinfo
  {author} {\bibfnamefont {S.}~\bibnamefont {Lamon}}, \bibinfo {author}
  {\bibfnamefont {B.}~\bibnamefont {Wang}}, \bibinfo {author} {\bibfnamefont
  {H.}~\bibnamefont {Ding}}, \bibinfo {author} {\bibfnamefont {J.}~\bibnamefont
  {Lin}}, \bibinfo {author} {\bibfnamefont {Q.}~\bibnamefont {Wang}}, \bibinfo
  {author} {\bibfnamefont {H.}~\bibnamefont {Luan}}, \bibinfo {author}
  {\bibfnamefont {M.}~\bibnamefont {Gu}},\ and\ \bibinfo {author}
  {\bibfnamefont {Q.}~\bibnamefont {Zhang}},\ }\bibfield  {title} {\bibinfo
  {title} {All-optical image transportation through a multimode fibre using a
  miniaturized diffractive neural network on the distal facet},\ }\href@noop {}
  {\bibfield  {journal} {\bibinfo  {journal} {Nature Photonics}\ ,\ \bibinfo
  {pages} {486}} (\bibinfo {year} {2025})}\BibitemShut {NoStop}%
\bibitem [{\citenamefont {Mididoddi}\ \emph {et~al.}(2025)\citenamefont
  {Mididoddi}, \citenamefont {Sharp}, \citenamefont {del Hougne}, \citenamefont
  {Horsley},\ and\ \citenamefont {Phillips}}]{mididoddi2025threading}%
  \BibitemOpen
  \bibfield  {author} {\bibinfo {author} {\bibfnamefont {C.~K.}\ \bibnamefont
  {Mididoddi}}, \bibinfo {author} {\bibfnamefont {C.}~\bibnamefont {Sharp}},
  \bibinfo {author} {\bibfnamefont {P.}~\bibnamefont {del Hougne}}, \bibinfo
  {author} {\bibfnamefont {S.~A.}\ \bibnamefont {Horsley}},\ and\ \bibinfo
  {author} {\bibfnamefont {D.~B.}\ \bibnamefont {Phillips}},\ }\bibfield
  {title} {\bibinfo {title} {Threading light through dynamic complex media},\
  }\href@noop {} {\bibfield  {journal} {\bibinfo  {journal} {Nature Photonics}\
  } (\bibinfo {year} {2025})}\BibitemShut {NoStop}%
\bibitem [{\citenamefont {Goodman}(2005)}]{goodman2005introduction}%
  \BibitemOpen
  \bibfield  {author} {\bibinfo {author} {\bibfnamefont {J.~W.}\ \bibnamefont
  {Goodman}},\ }\href@noop {} {\emph {\bibinfo {title} {Introduction to Fourier
  optics}}}\ (\bibinfo  {publisher} {Roberts and Company publishers},\ \bibinfo
  {address} {Greenwood Village, CO},\ \bibinfo {year} {2005})\BibitemShut
  {NoStop}%
\bibitem [{\citenamefont {Benesty}\ \emph {et~al.}(2009)\citenamefont
  {Benesty}, \citenamefont {Chen}, \citenamefont {Huang},\ and\ \citenamefont
  {Cohen}}]{Benesty2009}%
  \BibitemOpen
  \bibfield  {author} {\bibinfo {author} {\bibfnamefont {J.}~\bibnamefont
  {Benesty}}, \bibinfo {author} {\bibfnamefont {J.}~\bibnamefont {Chen}},
  \bibinfo {author} {\bibfnamefont {Y.}~\bibnamefont {Huang}},\ and\ \bibinfo
  {author} {\bibfnamefont {I.}~\bibnamefont {Cohen}},\ }\bibinfo {title}
  {Pearson correlation coefficient},\ in\ \href@noop {} {\emph {\bibinfo
  {booktitle} {Noise Reduction in Speech Processing}}}\ (\bibinfo  {publisher}
  {Springer Berlin Heidelberg},\ \bibinfo {address} {Berlin, Heidelberg},\
  \bibinfo {year} {2009})\ pp.\ \bibinfo {pages} {1--4}\BibitemShut {NoStop}%
\bibitem [{\citenamefont {LeCun}\ \emph {et~al.}(1998)\citenamefont {LeCun},
  \citenamefont {Bottou}, \citenamefont {Bengio},\ and\ \citenamefont
  {Haffner}}]{lecun1998gradient}%
  \BibitemOpen
  \bibfield  {author} {\bibinfo {author} {\bibfnamefont {Y.}~\bibnamefont
  {LeCun}}, \bibinfo {author} {\bibfnamefont {L.}~\bibnamefont {Bottou}},
  \bibinfo {author} {\bibfnamefont {Y.}~\bibnamefont {Bengio}},\ and\ \bibinfo
  {author} {\bibfnamefont {P.}~\bibnamefont {Haffner}},\ }\bibfield  {title}
  {\bibinfo {title} {Gradient-based learning applied to document recognition},\
  }\href@noop {} {\bibfield  {journal} {\bibinfo  {journal} {Proceedings of the
  IEEE}\ }\textbf {\bibinfo {volume} {86}},\ \bibinfo {pages} {2278} (\bibinfo
  {year} {1998})}\BibitemShut {NoStop}%
\bibitem [{\citenamefont {Kupianskyi}\ \emph {et~al.}(2024)\citenamefont
  {Kupianskyi}, \citenamefont {Horsley},\ and\ \citenamefont
  {Phillips}}]{kupianskyi2024all}%
  \BibitemOpen
  \bibfield  {author} {\bibinfo {author} {\bibfnamefont {H.}~\bibnamefont
  {Kupianskyi}}, \bibinfo {author} {\bibfnamefont {S.~A.}\ \bibnamefont
  {Horsley}},\ and\ \bibinfo {author} {\bibfnamefont {D.~B.}\ \bibnamefont
  {Phillips}},\ }\bibfield  {title} {\bibinfo {title} {All-optically untangling
  light propagation through multimode fibers},\ }\href@noop {} {\bibfield
  {journal} {\bibinfo  {journal} {Optica}\ }\textbf {\bibinfo {volume} {11}},\
  \bibinfo {pages} {101} (\bibinfo {year} {2024})}\BibitemShut {NoStop}%
\bibitem [{\citenamefont {B{\=u}tait{\.e}}\ \emph {et~al.}(2022)\citenamefont
  {B{\=u}tait{\.e}}, \citenamefont {Kupianskyi}, \citenamefont
  {{\v{C}}i{\v{z}}m{\'a}r},\ and\ \citenamefont {Phillips}}]{butaite2022build}%
  \BibitemOpen
  \bibfield  {author} {\bibinfo {author} {\bibfnamefont {U.~G.}\ \bibnamefont
  {B{\=u}tait{\.e}}}, \bibinfo {author} {\bibfnamefont {H.}~\bibnamefont
  {Kupianskyi}}, \bibinfo {author} {\bibfnamefont {T.}~\bibnamefont
  {{\v{C}}i{\v{z}}m{\'a}r}},\ and\ \bibinfo {author} {\bibfnamefont {D.~B.}\
  \bibnamefont {Phillips}},\ }\bibfield  {title} {\bibinfo {title} {How to
  build the “optical inverse” of a multimode fibre},\ }\href@noop {}
  {\bibfield  {journal} {\bibinfo  {journal} {Intelligent Computing}\ }
  (\bibinfo {year} {2022})}\BibitemShut {NoStop}%
\bibitem [{\citenamefont {Chen}\ \emph {et~al.}(2020)\citenamefont {Chen},
  \citenamefont {Li}, \citenamefont {Zhang}, \citenamefont {Wang},\ and\
  \citenamefont {Chen}}]{aircraft2020}%
  \BibitemOpen
  \bibfield  {author} {\bibinfo {author} {\bibfnamefont {J.~Y.}\ \bibnamefont
  {Chen}}, \bibinfo {author} {\bibfnamefont {H.~W.}\ \bibnamefont {Li}},
  \bibinfo {author} {\bibfnamefont {G.}~\bibnamefont {Zhang}}, \bibinfo
  {author} {\bibfnamefont {S.}~\bibnamefont {Wang}},\ and\ \bibinfo {author}
  {\bibfnamefont {T.~Q.}\ \bibnamefont {Chen}},\ }\bibfield  {title} {\bibinfo
  {title} {Dataset of aircraft classification in remote sensing images},\
  }\href@noop {} {\bibfield  {journal} {\bibinfo  {journal} {Glob. Change Data
  Discov}\ }\textbf {\bibinfo {volume} {2}},\ \bibinfo {pages} {183} (\bibinfo
  {year} {2020})}\BibitemShut {NoStop}%
\bibitem [{\citenamefont {Krizhevsky}\ and\ \citenamefont
  {Hinton}(2009)}]{Krizhevsky2009LearningML}%
  \BibitemOpen
  \bibfield  {author} {\bibinfo {author} {\bibfnamefont {A.}~\bibnamefont
  {Krizhevsky}}\ and\ \bibinfo {author} {\bibfnamefont {G.}~\bibnamefont
  {Hinton}},\ }\bibfield  {title} {\bibinfo {title} {Learning multiple layers
  of features from tiny images},\ }\href
  {https://www.cs.toronto.edu/~kriz/learning-features-2009-TR.pdf} {\bibfield
  {journal} {\bibinfo  {journal} {Technical Report}\ } (\bibinfo {year}
  {2009})}\BibitemShut {NoStop}%
\bibitem [{\citenamefont {Wang}\ \emph {et~al.}(2023)\citenamefont {Wang},
  \citenamefont {Sohoni}, \citenamefont {Wright}, \citenamefont {Stein},
  \citenamefont {Ma}, \citenamefont {Onodera}, \citenamefont {Anderson},\ and\
  \citenamefont {McMahon}}]{wang2023image}%
  \BibitemOpen
  \bibfield  {author} {\bibinfo {author} {\bibfnamefont {T.}~\bibnamefont
  {Wang}}, \bibinfo {author} {\bibfnamefont {M.~M.}\ \bibnamefont {Sohoni}},
  \bibinfo {author} {\bibfnamefont {L.~G.}\ \bibnamefont {Wright}}, \bibinfo
  {author} {\bibfnamefont {M.~M.}\ \bibnamefont {Stein}}, \bibinfo {author}
  {\bibfnamefont {S.-Y.}\ \bibnamefont {Ma}}, \bibinfo {author} {\bibfnamefont
  {T.}~\bibnamefont {Onodera}}, \bibinfo {author} {\bibfnamefont {M.~G.}\
  \bibnamefont {Anderson}},\ and\ \bibinfo {author} {\bibfnamefont {P.~L.}\
  \bibnamefont {McMahon}},\ }\bibfield  {title} {\bibinfo {title} {Image
  sensing with multilayer nonlinear optical neural networks},\ }\href@noop {}
  {\bibfield  {journal} {\bibinfo  {journal} {Nature Photonics}\ }\textbf
  {\bibinfo {volume} {17}},\ \bibinfo {pages} {408} (\bibinfo {year}
  {2023})}\BibitemShut {NoStop}%
\bibitem [{\citenamefont {Meglinski}\ \emph {et~al.}(2024)\citenamefont
  {Meglinski}, \citenamefont {Lopushenko}, \citenamefont {Sdobnov},\ and\
  \citenamefont {Bykov}}]{Meglinski2024}%
  \BibitemOpen
  \bibfield  {author} {\bibinfo {author} {\bibfnamefont {I.}~\bibnamefont
  {Meglinski}}, \bibinfo {author} {\bibfnamefont {I.}~\bibnamefont
  {Lopushenko}}, \bibinfo {author} {\bibfnamefont {A.}~\bibnamefont
  {Sdobnov}},\ and\ \bibinfo {author} {\bibfnamefont {A.}~\bibnamefont
  {Bykov}},\ }\bibfield  {title} {\bibinfo {title} {Phase preservation of
  orbital angular momentum of light in multiple scattering environment},\
  }\href@noop {} {\bibfield  {journal} {\bibinfo  {journal} {Light: Science \&
  Applications}\ }\textbf {\bibinfo {volume} {13}},\ \bibinfo {pages} {214}
  (\bibinfo {year} {2024})}\BibitemShut {NoStop}%
\bibitem [{\citenamefont {Khanom}\ \emph {et~al.}(2024)\citenamefont {Khanom},
  \citenamefont {Mohamed}, \citenamefont {Lopushenko}, \citenamefont {Sdobnov},
  \citenamefont {Doronin}, \citenamefont {Bykov}, \citenamefont {Rafailov},\
  and\ \citenamefont {Meglinski}}]{khanom2024twists}%
  \BibitemOpen
  \bibfield  {author} {\bibinfo {author} {\bibfnamefont {F.}~\bibnamefont
  {Khanom}}, \bibinfo {author} {\bibfnamefont {N.}~\bibnamefont {Mohamed}},
  \bibinfo {author} {\bibfnamefont {I.}~\bibnamefont {Lopushenko}}, \bibinfo
  {author} {\bibfnamefont {A.}~\bibnamefont {Sdobnov}}, \bibinfo {author}
  {\bibfnamefont {A.}~\bibnamefont {Doronin}}, \bibinfo {author} {\bibfnamefont
  {A.}~\bibnamefont {Bykov}}, \bibinfo {author} {\bibfnamefont
  {E.}~\bibnamefont {Rafailov}},\ and\ \bibinfo {author} {\bibfnamefont
  {I.}~\bibnamefont {Meglinski}},\ }\bibfield  {title} {\bibinfo {title}
  {Twists through turbidity: propagation of light carrying orbital angular
  momentum through a complex scattering medium},\ }\href@noop {} {\bibfield
  {journal} {\bibinfo  {journal} {Scientific Reports}\ }\textbf {\bibinfo
  {volume} {14}},\ \bibinfo {pages} {20662} (\bibinfo {year}
  {2024})}\BibitemShut {NoStop}%
\bibitem [{\citenamefont {Goodman}(2007)}]{goodman2007speckle}%
  \BibitemOpen
  \bibfield  {author} {\bibinfo {author} {\bibfnamefont {J.~W.}\ \bibnamefont
  {Goodman}},\ }\href@noop {} {\emph {\bibinfo {title} {Speckle phenomena in
  optics: theory and applications}}}\ (\bibinfo  {publisher} {Roberts and
  Company Publishers},\ \bibinfo {address} {Greenwood Village, CO},\ \bibinfo
  {year} {2007})\BibitemShut {NoStop}%
\bibitem [{\citenamefont {Karniadakis}\ \emph {et~al.}(2021)\citenamefont
  {Karniadakis}, \citenamefont {Kevrekidis}, \citenamefont {Lu}, \citenamefont
  {Perdikaris}, \citenamefont {Wang},\ and\ \citenamefont
  {Yang}}]{karniadakis2021physics}%
  \BibitemOpen
  \bibfield  {author} {\bibinfo {author} {\bibfnamefont {G.~E.}\ \bibnamefont
  {Karniadakis}}, \bibinfo {author} {\bibfnamefont {I.~G.}\ \bibnamefont
  {Kevrekidis}}, \bibinfo {author} {\bibfnamefont {L.}~\bibnamefont {Lu}},
  \bibinfo {author} {\bibfnamefont {P.}~\bibnamefont {Perdikaris}}, \bibinfo
  {author} {\bibfnamefont {S.}~\bibnamefont {Wang}},\ and\ \bibinfo {author}
  {\bibfnamefont {L.}~\bibnamefont {Yang}},\ }\bibfield  {title} {\bibinfo
  {title} {Physics-informed machine learning},\ }\href@noop {} {\bibfield
  {journal} {\bibinfo  {journal} {Nature Reviews Physics}\ }\textbf {\bibinfo
  {volume} {3}},\ \bibinfo {pages} {422} (\bibinfo {year} {2021})}\BibitemShut
  {NoStop}%
\end{thebibliography}%


\subsection*{Data availability}
The code and the data will be available at \url{https://github.com/FishWoWater/ScatteringODNN} when the paper is officially published. 


\subsection*{Acknowledgements}
We would like to thank Professor Xin Yuan of Westlake University for his invaluable suggestions, and Professor Honglin Liu of the Shanghai Institute of Optics and Fine Mechanics for her precious advice as well. This work was supported by the National Natural Science Foundation of China (No. 62401359), the fund of the State Key Laboratory of Photonics and Communications, the Innovation Program for Quantum Science and Technology (Grant No. 2021ZD0300703), Shanghai Municipal Science and Technology Major Project (2019SHZDZX01) and SJTU-Lenovo Collaboration Project (202407SJTU01-LR019).

\subsection*{Author contributions} 
G.Z. conceived the research project. T.X. designed the scheme.
Y.L. constructed the theoretical model, operated the numerical simulation with assistance from J.W., M.H. and J.F., carried out the experiment with assistance from J.H. and Z.Z..
All the authors discussed and contributed to the writing of the manuscript.

\subsection*{Competing interests} 
The authors declare no competing interests.

\end{document}